\documentclass[12pt]{article}

\usepackage[dvips]{graphicx}
\usepackage{amssymb}

\makeatletter
 \@addtoreset{equation}{section}
 
\makeatother

\begin{document}


\begin{titlepage}

\renewcommand{\thefootnote}{\fnsymbol{footnote}}

\begin{flushright}
\begin{tabular}{l}
UTHEP-590\\
RIKEN-TH-168\\
\end{tabular}
\end{flushright}

\bigskip

\begin{center}
{\Large \bf Light-Cone Gauge String Field Theory\\
 in Noncritical Dimensions\\
}
\end{center}

\bigskip

\begin{center}
{\large
Yutaka Baba}${}^{a}$\footnote{e-mail:
        ybaba@riken.jp},
{\large Nobuyuki Ishibashi}${}^{b}$\footnote{e-mail:
        ishibash@het.ph.tsukuba.ac.jp},
{\large Koichi Murakami}${}^{a}$\footnote{e-mail:
        murakami@riken.jp}
\end{center}

\begin{center}
${}^{a}${\it Theoretical Physics Laboratory, RIKEN,\\
         Wako, Saitama 351-0198, Japan}
\end{center}
\begin{center}
${}^{b}${\it
Institute of Physics, University of Tsukuba,\\
Tsukuba, Ibaraki 305-8571, Japan}\\
\end{center}

\bigskip

\bigskip

\bigskip

\begin{abstract}
We study light-cone gauge string field theory 
in noncritical space-time dimensions. 
Such a theory corresponds to a string theory in 
a Lorentz noninvariant background. 
We identify the worldsheet theory for the longitudinal 
coordinate variables $X^\pm$ and study its properties. 
It is a CFT with the right value of Virasoro central charge,
using which we propose a BRST invariant formulation 
of the worldsheet theory. 
\end{abstract}

\setcounter{footnote}{0}
\renewcommand{\thefootnote}{\arabic{footnote}}

\end{titlepage}

\section{Introduction}
Light-cone gauge string field theory provides a useful way to 
define string theories
\cite{Kaku:1974zz}\cite{Mandelstam:1973jk}\cite{Cremmer:1974ej}. 
Given the action, it is possible to 
define the amplitudes and calculate them perturbatively, 
although we should check if they are well-defined. 
Since it is a gauge fixed theory, 
there is no problem in considering a light-cone gauge string 
field theory in noncritical space-time dimensions. 
Being noncritical, it does not possess the space-time 
Lorentz invariance. 
It should correspond to a string theory in a Lorentz noninvariant 
background. 
In other words, 
we should be able to find a BRST invariant worldsheet theory, 
with a nonstandard $X^\pm$ part. 

What we would like to do in this paper is to study this $X^\pm$ theory. 
We give the energy-momentum tensor and the action of the theory 
and calculate the correlation functions. 
We show that the energy-momentum tensor possesses 
desired properties, and construct a BRST invariant formulation 
of the worldsheet theory.  

The reason why we are interested in this theory is 
that it can be used to regularize the string field theory. 
The dimensional regularization is one of the most powerful 
regularizations for ordinary quantum field theory 
and it may be useful also in string theory. 
In particular, 
in the light-cone gauge superstring field theory,
unwanted divergences occur even at the tree level, 
because of the transverse supercurrent insertions 
at the interaction points of the vertices. 
In Ref.~\cite{Baba:2009kr},
we have proposed a dimensional regularization scheme
to deal with these divergences. 
We have found that the divergences of the tree amplitudes
can be regularized by shifting the number of the space-time dimensions. 
We have checked that 
the results of the first quantized formalism are reproduced
without any counterterms for the four-point case.
In order to proceed further we need to show that 
the dimensional regularization preserves important symmetries 
of the theory. 
If the light-cone gauge string field theory 
corresponds to a BRST invariant formulation 
even in noncritical dimensions, 
it means that the dimensional regularization preserves the BRST 
symmetry. 

In this paper, we deal with the case of closed bosonic string field theory. 
We restrict ourselves to the tree amplitude and 
consider the worldsheet theory on the complex plane. 
The organization of this paper is as follows.
In section \ref{sec:LCcov}, 
we present a way to rewrite light-cone gauge 
string amplitudes in a BRST invariant form. 
We follow the procedure in noncritical dimensions, 
and show what kind of worldsheet theory of $X^\pm$ 
should appear in the end. 
In section \ref{sec:XpmCFT}, we study the theory for  $X^{\pm}$ 
and show that it is a CFT with the right Virasoro central charge. 
Namely, combining the CFT for $X^{\pm}$ constructed here
with the worldsheet theory for the light-cone gauge
strings in noncritical dimensions,
we obtain a CFT with the central charge $26$
by which we can define a BRST invariant worldsheet theory.
In section \ref{sec:BRST}, 
we show that the tree amplitudes in the noncritical case 
can be written in a BRST invariant form.
Section \ref{sec:discussions} is devoted to discussions.
In appendix \ref{sec:action},
we present the action of the light-cone gauge
bosonic string field theory
in $d$ dimensions.
In appendix \ref{sec:Mandelstam},
the Mandelstam mapping is given.
In appendix \ref{sec:Gamma},
we present a way to calculate $\Gamma [\phi ]$, 
which is used in sections \ref{sec:LCcov} and \ref{sec:XpmCFT}.
In appendix \ref{sec:relations},
the details of the calculations 
in section \ref{sec:XpmCFT} are presented.

\section{Relation between Light-Cone Gauge Amplitudes 
         and Covariant Ones}
\label{sec:LCcov}
\subsection{Critical case}

In order to study the noncritical case, it is useful to consider
the relation between light-cone gauge amplitudes and covariant ones
for critical strings. 
One can calculate the amplitudes starting from 
the light-cone gauge string field theory action given in appendix 
\ref{sec:action}. 
The tree amplitudes can be expressed by path integral 
on the light-cone diagrams, on which the complex $\rho$ coordinate
is introduced as usual.
Via the Mandelstam mapping $\rho (z)$, 
which is given in appendix \ref{sec:Mandelstam}, 
one can express them using 
correlation functions of vertex operators
on the $z$-plane endowed with the metric
\begin{equation}
ds^{2} = 
 d\rho d\bar{\rho} =  e^{\phi}dzd\bar{z}\;,
\qquad
\phi = \ln \left( \partial \rho \bar{\partial} \bar{\rho}
           \right)~.
 \label{eq:metric}
\end{equation}
The correlation functions we should consider are
\begin{equation}
F
=
(2\pi)^{2} \delta\biggl( \sum_{r=1}^{N} p^{+}_{r} \biggr)
\delta \biggl( \sum_{r=1}^{N} p^{-}_{r} \biggr)
\int
\left[dX^{i}\right]_{\phi}  e^{-S_{X^{i}}}
\prod_{r=1}^{N}V_{r}^{\mathrm{LC}}\;.
\label{eq:correlation}
\end{equation}
Here $S_{X^{i}}$ denotes the free action for
the $24$ transverse coordinates 
$X^i$ and 
$V_{r}^{\mathrm{LC}}$ denotes the vertex operator. 
We assume that the $r$-th external state is of the form 
\begin{equation}
\alpha^{i_{1}(r)}_{-n_{1}} \cdots
\tilde{\alpha}^{\tilde{\imath}_{1}(r)}_{-\tilde{n}_{1}} \cdots
 |p^-,p^{i} \rangle_r~,
\label{eq:stringstate-LC}
\end{equation}
and the vertex operator should be
\begin{equation}
V^{\mathrm{LC}}_{r} 
\sim
\left.
\alpha_r
 \, \frac{i \partial^{n_{1}} X^{i_{1} (r)} (w_{r})}
       {(n_{1}-1)!}
  \cdots
  \frac{i \bar{\partial}^{\tilde{n}_{1}}
          X^{\tilde{\imath}_{1} (r)} (\bar{w}_{r})}
        {(\tilde{n}_{1} -1)!}
  \cdots
 e^{i p^{i}_{r} X^{i}-p_r^- \tau_{0}^{(r)}}(w_r,\bar{w}_r)
\right|_{w_r=\bar{w}_r=0}  ~,
\label{eq:VrLC}
\end{equation}
where $\tau_{0}^{(r)} $
is defined in eq.(\ref{eq:rho-zIr}),
$w_{r}$ is the coordinate of the unit disk of the $r$-th
external string given in eq.(\ref{eq:rho-wr})
and $\alpha_{r}=2p^{+}_{r}$.
The on-shell and the level-matching conditions
require that
\begin{equation}
\frac{1}{2} \left( -2p^{+}_{r}p^{-}_{r}+p^{i}_{r}p^{i}_{r}
            \right)
 + \mathcal{N}_{r} = 1~,
\qquad
\mathcal{N}_{r} \equiv \sum_{i} n_{i} =\sum_{j} \tilde{n}_{j}~.
\label{eq:on-shell-cov}
\end{equation}

The path integral measure $\left[dX^{i}\right]_{\phi}$ should be
defined using the metric (\ref{eq:metric}). It is related to
the measure $\left[dX^{i}\right]$ which is defined with the flat
metric $ds^{2}=dzd\bar{z}$ as
\begin{eqnarray*}
\left[dX^{i}\right]_{\phi} 
& \propto & 
\left[dX^{i}\right]e^{-\Gamma\left[\phi\right]}\;,
\end{eqnarray*}
Roughly speaking, $\Gamma\left[\phi\right]$ is given by the Liouville
action
\begin{equation}
\Gamma\left[\phi\right] 
 \sim 
-\frac{24}{24\pi}\int d^{2}z
\,  \partial \phi \bar{\partial}\phi ~,
\label{eq:liouville}
\end{equation}
where $d^{2}z=d(\mathop{\mathrm{Re}}z)d(\mathop{\mathrm{Im}}z)$.
Since $\phi$ diverges at the poles and zeros of $\partial\rho$
by the definition (\ref{eq:metric}), the Liouville
action is not well-defined at these points. 
$\Gamma[\phi]$ can be evaluated
by regularizing the divergences and carefully taking various effects
into account \cite{Mandelstam:1985ww}\cite{Green:1987mn}. 
We present an alternative derivation of $\Gamma[\phi]$ 
in appendix \ref{sec:Gamma} and give an explicit form in 
eq.(\ref{eq:Gamma-c}).\footnote{
As is mentioned in appendix \ref{sec:Gamma}, 
this form of $\Gamma[\phi]$ is only up to a constant which can be
fixed by factorization. 
With the string field action in eq.(\ref{eq:SFTaction}) 
and the $\Gamma[\phi]$ in eq.(\ref{eq:Gamma-c}), 
one can see that the factor is fixed as \cite{Baba:2009kr}
\begin{eqnarray*}
\left[dX^{i}\right]_{\phi} 
& \sim & 
\left[dX^{i}\right]
\mathrm{sgn}\left(\prod_{r=1}^N\alpha_r\right)
e^{-\Gamma\left[\phi\right]}\;,
\end{eqnarray*}
up to a numerical factor. 
We ignore the
phase factor $\mathrm{sgn}\left(\prod_{r=1}^N\alpha_r\right)$
in the following, 
because it does not play any important roles.
}
Thus eq.(\ref{eq:correlation}) can be rewritten as
\begin{equation}
F
\sim (2\pi)^{2} \delta\biggl( \sum_{r=1}^{N} p^{+}_{r} \biggr)
  \delta \biggl( \sum_{r=1}^{N} p^{-}_{r} \biggr)
\int\left[dX^{i}\right]
e^{-S_{X^{i}}-\Gamma[\phi]}
\prod_{r}V_{r}^{\mathrm{LC}}\;.
\label{eq:correlation2}
\end{equation}

\subsubsection*{Longitudinal coordinates}

In order to covariantize the amplitudes, we need to introduce the
longitudinal coordinates. 
The light-cone gauge condition implies that 
$X^+$ equals the Lorentzian time on the light-cone diagram, 
which means on the $z$-plane, 
\begin{equation}
X^+(z,\bar{z})
=
-\frac{i}{2}\left(\rho (z)+\bar{\rho}(\bar{z})\right)~.
\end{equation}
Therefore we introduce the variable $X^+$ with
the delta function 
$\delta \left(X^+ + \frac{i}{2}(\rho +\bar{\rho})\right)$, 
which can be expressed as
\begin{eqnarray}
\delta\left(X^{+}+\frac{i}{2}\left(\rho+\bar{\rho}\right)\right)
&\sim&
\int
\left[dX^{\prime -}
\right]
e^{-\frac{1}{\pi}\int d^{2}z
 X^{\prime -}\partial\bar{\partial}X^+}
\prod_{r=1}^{N}
e^{-ip_{r}^{+}X^{\prime-}}\left(Z_{r},\bar{Z}_{r}\right)
\nonumber
\\
& &
\hspace{5mm}
\times
\int
\left[
db^{\prime}dc^{\prime}d\tilde{b}^{\prime}d\tilde{c}^{\prime}
\right]
e^{-\frac{1}{\pi}\int d^{2}z
   \left( b^{\prime}\bar{\partial}c^{\prime}
          +\tilde{b}^{\prime} \partial \tilde{c}^{\prime}\right)}
c'(\infty )\tilde{c}'(\infty )~. 
 \label{eq:X+delta}
\end{eqnarray}
Eq.(\ref{eq:X+delta}) should be considered as the formal Euclideanized 
version of a Lorentzian path integral.\footnote{
In the light-cone gauge, the worldsheet should be inherently 
with Lorentzian signature. In this paper, we use the Euclideanized 
expressions, which look more familiar. 
}
The Grassmann odd fields 
$b^{\prime},c^{\prime},\tilde{b}^{\prime},\tilde{c}^{\prime}$
of conformal weights
$\left(1,0\right)$, $ \left(0,0\right)$, $\left(0,1\right)$,
$\left(0,0\right)$
are introduced to cancel the determinant factor 
$\left(\det \partial\bar{\partial}\right)^{-1}$.

With these variables, we can rewrite the right hand side of 
eq.(\ref{eq:correlation2}) as 
\begin{eqnarray}
F
&\sim&
 \frac{2\pi \delta \left( \sum_{r=1}^{N} p^{-}_{r} \right)}
      {2 \pi \delta (0)}
\int
\left[dX^+dX^{\prime -}dX^{i}
db^{\prime}dc^{\prime}d\tilde{b}^{\prime}d\tilde{c}^{\prime}
\right]
e^{-S_{X^{i}}-S_\pm^\prime -S_{b^{\prime}c^{\prime}}}
\nonumber\\
& &
\hspace{10em}
\times
c^\prime (\infty )\tilde{c}^\prime (\infty )
\prod_{r=1}^{N}
 \left(V_{r}^{\mathrm{LC}}e^{-ip_{r}^{+}X^{\prime-}}
\left(Z_{r},\bar{Z}_{r}\right)
\right)
\;,
\label{eq:correlation3}
\end{eqnarray}
where
\begin{eqnarray}
S_{b^{\prime}c^{\prime}} 
& = & 
\frac{1}{\pi}\int d^{2}z
\left(
b^{\prime}\bar{\partial}c^{\prime}
 +\tilde{b}^{\prime}\partial\tilde{c}^{\prime}
\right)\;,
\nonumber \\
S^\prime_{\pm} 
& = & 
-\frac{1}{2\pi}\int d^{2}z
\left( \partial X^{+}\bar{\partial}X^{\prime -}
      +\bar{\partial} X^{+} \partial X^{\prime -}
\right)
+\Gamma\left[
\ln\left(-4\partial X^{+}\bar{\partial}X^{+}\right)
\right]
\;.
\label{eq:Xprimeaction}
\end{eqnarray}
We consider $S_{X^i}+S^\prime_\pm +S_{b^{\prime}c^{\prime}}$ as the 
worldsheet action for these variables. 
The action $S^\prime_\pm$ is with the interaction term 
which depends only on $X^+$. 
This interaction term does not affect the correlation functions 
with less than two insertions of $X^{\prime-}$, 
and they coincide with those of the free theory.  
Thus one can show the OPE
\begin{eqnarray}
X^+(z,\bar{z})X^+(z',\bar{z}')
&\sim&
\mathrm{regular}~,
\nonumber
\\
X^+(z,\bar{z})X^{\prime-}(z',\bar{z}')
&\sim&
\ln \left|z-z'\right|^2~.
\end{eqnarray}

{}From the worldsheet action, 
the energy-momentum tensor can be obtained as 
\begin{equation}
T\left(z\right)
=
\partial X^{+}\partial X^{\prime-}
-2\left\{ X^{+},z\right\} 
-\frac{1}{2}\partial X^{i}\partial X^{i}
-b^{\prime}\partial c^{\prime}
\;,
\label{eq:emtensor}
\end{equation}
when $z\neq Z_{r},z_{I}$. 
Here 
\begin{eqnarray*}
\left\{ X^{+},z\right\}  
& = & 
\frac{\partial^{3}X^{+}}{\partial X^{+}}
-\frac{3}{2}\left(\frac{\partial^{2}X^{+}}{\partial X^{+}}\right)^{2}
\\
& = & 
-\frac{1}{2}
\left(\partial
\left(
\ln\left(-4\partial X^{+}\bar{\partial}X^{+}\right)
\right)
\right)^{2}
+
\partial^{2}\left(
\ln\left(-4\partial X^{+}\bar{\partial}X^{+}\right)
\right)\;,
\end{eqnarray*}
denotes the Schwarzian derivative. This term can be derived from the
variation of $\Gamma$ which coincides with the Liouville action except
for the singular points.

\subsubsection*{Covariant variables}

The covariant expression for the amplitudes can be obtained 
by introducing
\cite{D'Hoker:1987pr}
\begin{eqnarray}
b & \equiv & \partial X^{+}b^{\prime}\;,\nonumber\\
c & \equiv & \left(\partial X^{+}\right)^{-1}c^{\prime}\;,
\label{eq:defbc}
\end{eqnarray}
and their anti-holomorphic counterparts $\tilde{b}$ and $\tilde{c}$. 
The fields $b$ and $c$ have now weights $\left(2,0\right)$
and $\left(-1,0\right)$ respectively.
We should also introduce 
\begin{equation}
X^{-}  \equiv  
X^{\prime-}+\frac{b^{\prime}c^{\prime}}{\partial X^{+}}
-\frac{\partial^{2}X^{+}}{2\left(\partial X^{+}\right)^{2}}
+\frac{\tilde{b}'\tilde{c}'}{\bar{\partial} X^{+}}
-\frac{\bar{\partial}^{2} X^{+}}
      {2 \left(\bar{\partial} X^{+} \right)^{2}}\;,
\label{eq:newvariables}
\end{equation}
so that the OPE's between $X^{-}$ and the ghosts are regular
and the energy-momentum tensor (\ref{eq:emtensor}) takes the form
\begin{eqnarray}
T (z) 
& = & \partial X^{+}\partial X^{-}
-\frac{1}{2}\partial X^{i}\partial X^{i}-2b\partial c-\partial bc\;,
\label{eq:emtensor2}
\end{eqnarray}
which coincides with the energy-momentum tensor in the conformal gauge. 
In the following, we would like to rewrite 
the correlation function (\ref{eq:correlation3}) using these new variables. 

\subsubsection*{DDF operators}
Let us first consider the vertex operator. 
{}From eq.(\ref{eq:emtensor2}), one can see that the action 
for the new variables should be the free action. 
$X^-$ appears essentially in the same way as $X^{\prime -}$ does 
in eq.(\ref{eq:X+delta}), and we obtain the delta function. 
Therefore $\rho (z)$ and $\bar{\rho}(\bar{z})$ 
which appear in the integrand 
can be replaced by $2iX^+_L(z)$ and $2iX^+_R(\bar{z})$, where 
$X^{+}_L(z)$ and $X^{+}_R(\bar{z})$ are the holomorphic and 
the anti-holomorphic part of $X^+$ respectively. 
We will denote the equality which holds under this
identification by $\approx$.
Thus the factor 
$\left.\frac{i\partial^n X^i(w_r)}{(n-1)!}\right|_{w_r=0}$ 
in the definition~(\ref{eq:VrLC}) of $V_r^\mathrm{LC}$
can be rewritten as
\begin{eqnarray}
\left.\frac{i\partial^n X^i(w_r)}{(n-1)!}\right|_{w_r=0}
&=&
\oint_0 \frac{dw_r}{2\pi i}i\partial X^i(w_r)w_r^{-n}
\nonumber
\\
&\approx&
\oint_{Z_r}\frac{dz}{2\pi i}i\partial X^i(z)
e^{-in\frac{X_L^+ (z)}{p^{+}_{r}}
   +n\frac{\tau_{0}^{(r)}+i\beta_r}{\alpha_r}}
\nonumber
\\
&=&
A_{-n}^{(r)i}e^{n\frac{\tau_0^{(r)}+i\beta_r}{\alpha_r}}~,
\end{eqnarray}
where $A_{-n}^{(r)i}$ is the DDF operator given by
\begin{equation}
A_{-n}^{(r)i}
\equiv
\oint_{Z_r}\frac{dz}{2\pi i}i\partial X^i(z)
e^{-in\frac{X_L^+ (z)}{p^{+}_{r}}}~.
\end{equation}
One can also use 
\begin{eqnarray}
e^{-i\left(p_{r}^{-} - \frac{2\mathcal{N}_{r}}{\alpha_r}
     \right) X^+}
\left(z,\bar{z}\right)
&\approx&
e^{
-\frac{1}{2}
\left(p_r^--\frac{2\mathcal{N}_{r}}{\alpha_r}\right)
\left(\rho +\bar{\rho}\right)}\left(z,\bar{z}\right)
\nonumber\\
&\sim&
\left|
z-Z_r
\right|^{-2\left(p_r^+p_r^--\mathcal{N}_{r}\right)}
e^{-2\left(p_r^+p_r^- - \mathcal{N}_{r}\right)
   \left(\frac{\tau_{0}^{(r)}}{\alpha_r}
   - \mathop{\mathrm{Re}}\bar{N}^{rr}_{00}\right)}
\end{eqnarray}
for $z\sim Z_r$, where
$\bar{N}^{rr}_{00}$ is defined in eq.(\ref{eq:Nbarrr00}). 
Using these, one can show that 
$V_r^\mathrm{LC}e^{-ip_r^+X^{\prime -}}(Z_r,\bar{Z}_r)$
in eq.(\ref{eq:correlation3})
subject to the on-shell condition (\ref{eq:on-shell-cov})
can be rewritten as
\begin{equation}
  V_r^\mathrm{LC}e^{-ip_r^+X^{\prime -}}(Z_r,\bar{Z}_r)
 \approx
\alpha_r
V_r^\mathrm{DDF}(Z_r,\bar{Z}_r)e^{2\mathrm{Re}\bar{N}^{rr}_{00}}~,
\label{eq:LCDDF}
\end{equation} 
where
\begin{equation}
V_r^\mathrm{DDF} (z,\bar{z})
=
A_{-n_1}^{(r)i_1}\cdots
\tilde{A}_{-\tilde{n}_1}^{(r)\tilde{\imath}_1}\cdots
\, :\! e^{-ip_r^+X^--i\left(p_r^--\frac{2\mathcal{N}_{r}}{\alpha_r}
                      \right)X^+ + ip_r^i X^i} (z,\bar{z})
    \! :
\label{eq:VDDF}
\end{equation}
is the vertex operator corresponding to the DDF state. 

\subsubsection*{Covariant expression}
One subtle point to notice in eq.(\ref{eq:defbc}) is that the variables $b,c$
should have zeros and poles at the zeros and poles of $\partial X^{+}$. 
Since the integration over $X^{-}$ leads to
the identification $X^{+} \approx - \frac{i}{2}(\rho +\bar{\rho})$,
the zeros and poles of $\partial X^{+}$ become
those of $\partial\rho$, namely
$z_{I}$ and $Z_{r}$. 
Accordingly $b$ should be inserted at $z_{I}$ and $c$ should 
be inserted at $Z_{r}$, 
if we wish to rewrite eq.(\ref{eq:correlation3})
\cite{D'Hoker:1987pr}\cite{Kunitomo:1987uf}
\cite{Sonoda:1987ra}\cite{D'Hoker:1989ae}. 
These insertions should come with appropriate factors made from 
$\rho ,\bar{\rho}$ or $X^+$
in order that eq.(\ref{eq:correlation3}) is reproduced.
Using eqs.(\ref{eq:LCDDF}) and (\ref{eq:exp-Gamma2}), 
one can show that eq.(\ref{eq:correlation3}) can be rewritten as
\begin{eqnarray}
 F
 &\sim& 
 \int\left[dX^{\mu}dbdcd\tilde{b}d\tilde{c}\right]
  e^{-S_{X}-S_{bc}}
\,   \left|\sum_{r}\alpha_{r}Z_{r}\right|^{2}
 \left(
 \lim_{z\to\infty}\frac{1}
 {\left|z\right|^{4}}
 c\left(z\right)\tilde{c}\left(\bar{z}\right)\right)
\nonumber \\
 &  & \hspace{1cm}
 \times\prod_{I}
 \left(
 \frac{b}{\partial^{2}\rho}(z_{I})
 \frac{\tilde{b}}{\bar{\partial}^{2}\bar{\rho}}(\bar{z}_{I})\right)
 \prod_{r=1}^{N} \left(c\tilde{c} V_{r}^{\mathrm{DDF}}\right)
          \left(Z_{r},\bar{Z}_{r}\right) \;,
 \label{eq:correlation4}
\end{eqnarray}
where $S_X$ and $S_{bc}$ are the free actions
for $X^\mu=(X^{\pm},X^{i})$
and for $b,c$ respectively. 

One can show that eq.(\ref{eq:correlation4}) 
yields the expression for the amplitudes in the 
conformal gauge. 
Using
\begin{eqnarray*}
\frac{b}{\partial^{2}\rho}(z_{I}) 
& = & 
\oint_{z_{I}}\frac{dz}{2\pi i}\frac{b}{\partial\rho}(z)
\;,\end{eqnarray*}
and deforming the contour, 
one can recast eq.(\ref{eq:correlation4}) into the form
\begin{eqnarray}
 F
 &\sim& 
 \int\left[dX^{\mu}dbdcd\tilde{b}d\tilde{c}\right]
   e^{-S_{X}-S_{bc}}\nonumber \\
 &  & \hspace{1cm}
 \times\prod_{I}
 \left(
   \oint_{C_{I}}\frac{dz}{2\pi i}\frac{b}{\partial\rho}(z)
   \oint_{C_{I}}
   \frac{d\bar{z}}{2\pi i}
     \frac{\tilde{b}}{\bar{\partial}\bar{\rho}} (\bar{z})
 \right)
 \prod_{r=1}^{N}
   \left(c\tilde{c}V_{r}^{\mathrm{DDF}}\right)
          \left(Z_{r},\bar{Z}_{r}\right)\;,
 \label{eq:correlation5}
\end{eqnarray}
where the integration contour $C_{I}$ is
depicted in Fig.~\ref{fig:Npt} in appendix \ref{sec:Gamma}
on the $\rho$-plane.

The amplitudes can be obtained by integrating 
the correlation function $F$ over the $N-3$ 
moduli parameters $\mathcal{T}_{I}$ defined 
in eq.(\ref{eq:mathcalT}) as
\begin{equation}
\mathcal{A}
\propto
\int \prod_I d^2\mathcal{T}_{I}
\, F(\mathcal{T}_I,\bar{\mathcal{T}}_I)~.
\label{eq:amplitude}
\end{equation}
One can see that the antighost insertion 
$\oint_{C_{I}}\frac{dz}{2\pi i}\frac{b}{\partial\rho}\left(z\right)$ 
corresponds to the quasiconformal vector field 
associated with the moduli parameter $\mathcal{T}_I$. 
The form of the amplitude is BRST invariant, because 
the vertex operator $c\tilde{c}V_r^\mathrm{DDF}$ is invariant and 
\begin{equation}
\left\{ 
Q_\mathrm{B},
\oint_{C_{I}}\frac{dz}{2\pi i}\frac{b}{\partial\rho}\left(z\right)
\right\}
=
\oint_{C_{I}}\frac{dz}{2\pi i}\frac{T}{\partial\rho}\left(z\right)~,
\end{equation}
yields a total derivative with respect to $\mathcal{T}_I$. 

We can change the integration variables to $Z_r~(r=3,4,\cdots ,N-1)$ 
and 
\begin{equation}
\mathcal{A}
\propto
\int \prod_{r=3}^{N-1} d^2Z_r
\left| 
\det
\left(
\frac{\partial \mathcal{T}_I}{\partial Z_r}
\right)
\right|^2
F(\mathcal{T}_I,\bar{\mathcal{T}}_I)~.
\end{equation}
With the expression (\ref{eq:correlation5}) for $F$, 
the determinant factor can be combined with the antighost factors as
\begin{eqnarray}
\det
\left(
\frac{\partial \mathcal{T}_I}{\partial Z_r}
\right)
\times\prod_{I}
\oint_{C_{I}}\frac{dz}{2\pi i}
\frac{b}{\partial\rho}\left(z\right)
&\propto&
\prod_{r=3}^{N-1}
\left(
\sum_I
\frac{\partial \mathcal{T}_I}{\partial Z_r}
\oint_{C_{I}}\frac{dz}{2\pi i}
\frac{b}{\partial\rho}\left(z\right)
\right)
\nonumber
\\
&=&
\prod_{r=3}^{N-1}
\left(
\sum_I
\oint_{C_{I}}
    \frac{dz}{2\pi i}
     \frac{\partial_{Z_{r}}
            \left( \rho (z) - \rho (z_{I}) \right)}
          {\partial \rho (z)}
     b(z)
     \right.
\nonumber
\\
& &
\hspace{1.5cm}
\left.
  - \sum_I\oint_{C_{I}}
    \frac{dz}{2\pi i}
     \frac{\partial_{Z_{r}}
            \left( \rho (z) - \rho (z_{I+1}) \right)}
          {\partial \rho (z)}
     b(z)
\right)
\nonumber
\\
&\propto&
\prod_{r=3}^{N-1}
\left(
\sum_{s=1}^N
\oint_{Z_s}
    \frac{dz}{2\pi i}
     \frac{\partial_{Z_r}
            \left( \rho (z) - \rho (z_{I^{(s)}}) \right)}
          {\partial \rho (z)}
     b(z)
\right)~.
\end{eqnarray}
Performing the contour integrals around $Z_s$ in the last line, 
we eventually obtain the familiar expression
\begin{eqnarray}
 \mathcal{A}
 &\propto& 
 \int\left[dX^{\mu}dbdcd\tilde{b}d\tilde{c}\right]
  e^{-S_{X}-S_{bc}}\nonumber \\
 &  & \hspace{1cm}
 \times
 \prod_{s=1,2,N}
 \left(c\tilde{c}V_{s}^{\mathrm{DDF}}\right)
    \left(Z_{s},\bar{Z}_{s}\right)
 \prod_{r=3}^{N-1}
 \int d^2Z_r
 V_{r}^{\mathrm{DDF}}\left(Z_{r},\bar{Z}_{r}\right)\;.
\end{eqnarray}

Hence the variables $X^{\pm},b,c$ can be identified with those in the 
covariant formulation.
{}From eq.(\ref{eq:newvariables}),
one can find the OPE's of the variables 
$X^{+},X^{\prime-},b^{\prime},c^{\prime}$
to be
\begin{eqnarray}
\partial X^{+}(z) \partial X^{+} (z')
  & \sim & \mathrm{regular}\;,
\nonumber \\
\partial X^{+} (z) \partial X^{\prime-}(z') 
  & \sim & \frac{1}{(z-z')^{2}}\;,
\nonumber \\
\partial X^{\prime-}(z) \partial X^{\prime-}(z')
 & \sim & -2\partial_{z} \partial_{z'}
         \left( \frac{1}{(z-z')^{2}}
                \frac{1}{\partial X^{+}(z)
                         \partial X^{+}(z')} \right)
\;,\nonumber \\
b'(z)c'(z')
  & \sim & \frac{1}{z-z'}\;,
\label{eq:OPE}
\end{eqnarray}
and regular otherwise.
These can also be deduced from the action (\ref{eq:Xprimeaction}).
Using these OPE's, one can easily show that 
the energy-momentum tensor (\ref{eq:emtensor}) satisfies
the Virasoro algebra of central charge $c=0$.

\subsection{Noncritical case}

The light-cone gauge string field theory in 
$d$ $(d \neq 26)$ space-time dimensions can be defined with 
the action given in appendix \ref{sec:action}. 
This time, the correlation functions we should consider 
is\footnote{
This time we have
\begin{eqnarray*}
\left[dX^{i}\right]_{\phi} 
& \sim & 
\left[dX^{i}\right]
\mathrm{sgn}\left(\prod_{r=1}^N\alpha_r\right)
e^{-\frac{d-2}{24}\Gamma\left[\phi\right]}\;,
\end{eqnarray*}
and we ignore the phase $\mathrm{sgn}\left(\prod_{r=1}^N\alpha_r\right)$ 
in the following.
}
\begin{equation}
F
\sim (2\pi)^{2} \delta\biggl( \sum_{r=1}^{N} p^{+}_{r} \biggr)
  \delta \biggl( \sum_{r=1}^{N} p^{-}_{r} \biggr)
\int\left[dX^{i}\right]
e^{-S_{X^{i}}-\frac{d-2}{24}\Gamma[\phi]}
\prod_{r}V_{r}^{\mathrm{LC}}\;.
\label{eq:noncriticalF}
\end{equation}
We can follow the above procedure and introduce the variables 
$X^\pm ,b,c$ 
without any problem.
In this case, eq.(\ref{eq:emtensor}) should be 
\begin{equation}
T(z)
=\partial X^{+}\partial X^{\prime-}
  - \frac{1}{2}\partial X^{i}\partial X^{i}
  -\frac{d-2}{12}\left\{ X^{+},z\right\}
  -b^{\prime}\partial c^{\prime}\;.
\label{eq:nenergymomentum}
\end{equation}
By using $X^{\pm},b,c$ defined in eq.(\ref{eq:newvariables}), 
$T (z)$ can be rewritten as 
\begin{equation}
T (z) 
 = \partial X^{+}\partial X^{-}
      - \frac{d-26}{12}\left\{ X^{+},z\right\}
      -\frac{1}{2}\partial X^{i}\partial X^{i}
      -2b\partial c - \partial bc\;,
\end{equation}
and the action $S_{X^{\pm}}$ should be 
\begin{equation}
S_{X^{\pm}} 
= 
-\frac{1}{2 \pi}\int d^{2}z
 \left( \partial X^{+} \bar{\partial} X^{-}
         + \bar{\partial} X^{+} \partial X^{-}
 \right)
+\frac{d-26}{24}\Gamma [\ln\left(-4\partial X^+\bar{\partial}X^+\right)]\;.
\label{eq:SXpm}
\end{equation}
Therefore in noncritical dimensions the worldsheet theory for $X^{\pm}$
is different from the usual free theory, 
and obviously the Lorentz symmetry is broken. 

What we should study is this theory for $X^{\pm}$. 
It is a conformal field theory similar
to the one for $X^{+}$ and $X^{\prime -}$ 
in the previous subsection. 
We will study its properties in the 
next section. 

Before closing this subsection, a comment is in order. 
For $d\neq 26$, one can define
\begin{eqnarray}
b^{\prime\prime}
&\equiv&
\left(\partial X^+\right)^\alpha b^\prime ~,
\nonumber
\\
c^{\prime\prime}
&\equiv&
\left(\partial X^+\right)^{-\alpha} c^\prime ~,
\nonumber
\\
X^{\prime\prime -}
&\equiv&
X^{\prime -}
+\alpha \frac{b'c'}{\partial X^+}
-\frac{\alpha}{2}
 \frac{\partial^2X^+}{(\partial X^+ )^2}
+\alpha \frac{\tilde{b}'\tilde{c}'}{\bar{\partial} X^{+}}
-\frac{\alpha}{2}
  \frac{\bar{\partial}^{2} X^{+}}{(\bar{\partial} X^{+})^{2}}~,
\end{eqnarray}
with
\begin{equation}
\alpha \left(\alpha +1\right)
=
\frac{d-2}{12}~.
\end{equation}
Using these variables, the energy-momentum tensor 
(\ref{eq:nenergymomentum}) can be written as
\begin{equation}
T\left(z\right)=\partial X^{+}\partial X^{\prime\prime -}-\frac{1}{2}\partial X^{i}\partial X^{i} 
-b^{\prime\prime}\partial c^{\prime\prime}
-\alpha \partial\left(b^{\prime\prime}c^{\prime\prime}\right)
\;.
\end{equation}
Therefore the worldsheet theory is a free theory with ghosts of 
noninteger spins. 
We may be able to study the theory using these variables, 
although we need to figure out the way to deal
with the ghost zero-modes.

\section{$X^{\pm}$ CFT}
\label{sec:XpmCFT}

\subsection{Action and correlation functions}
\label{sec:CFTaction}

The theory we would like to consider is with the action (\ref{eq:SXpm})
and the energy-momentum tensor
\begin{equation}
T_{X^\pm}(z)
=
\partial X^{+}\partial X^{-}-\frac{d-26}{12}\left\{ X^{+},z\right\}~.
\label{eq:emtensorXpm}
\end{equation}
This theory  will be well-defined
if $\partial X^{+}$ has a nonvanishing expectation value. 
As in the case of the theory for $X^{+}$ and $X^{\prime -}$,
we always consider this theory
in the presence of the insertions of the vertex operators
$e^{-ip^{+}_{r} X^{-}} (Z_{r},\bar{Z}_{r})$
($r=1,\ldots,N$) with $\sum_{r=1}^{N} p_{r}^{+}=0$,
so that the classical equation of motion implies
\begin{equation}
X^{+} (z,\bar{z}) 
=
-i\sum_{r=1}^{N} p_{r}^{+}\ln\left|z-Z_{r}\right|^{2}
 =  -\frac{i}{2}
       \left(\rho\left(z\right)
             +\bar{\rho}\left(\bar{z} \right)\right)~.
\label{eq:vevX+}
\end{equation}
Thus the quantities we would like to calculate are the 
expectation values
\begin{eqnarray}
\lefteqn{\left\langle 
F\left[X^{+},X^{-}\right]
\prod_{r=1}^{N}
        e^{-ip_{r}^{+}X^{-}}\left(Z_{r} ,\bar{Z}_{r}\right)
\right\rangle
} \nonumber\\
&&
\equiv
   \int\left[dX^{+}dX^{-}\right]
    e^{-S_{X^\pm}}F\left[X^{+},X^{-}\right]
    \prod_{r=1}^{N}
        e^{-ip_{r}^{+}X^{-}}\left(Z_{r} ,\bar{Z}_{r}\right),
\end{eqnarray}
for functionals $F\left[X^{+},X^{-}\right]$ which satisfy
\begin{equation}
F\left[X^{+}+\epsilon_+,X^{-}+\epsilon_-\right]
=
F\left[X^{+},X^{-}\right]~,
\end{equation}
for arbitrary constants $\epsilon_\pm$. 

For functionals $F\left[X^{+}\right]$ which do not depend on $X^-$, 
it is formally possible to perform the path integral
as in eq.(\ref{eq:X+delta}) and obtain
\begin{equation}
\left\langle 
F\left[X^{+}\right]
\prod_{r=1}^{N}
        e^{-ip_{r}^{+}X^{-}}\left(Z_{r} ,\bar{Z}_{r}\right)
\right\rangle
\sim
F\left[-\frac{i}{2}\left(\rho+\bar{\rho}\right)\right]
\exp
\left(-\frac{d-26}{24}\Gamma 
  \left[\ln\left(\partial \rho \bar{\partial}\bar{\rho}\right)\right]
\right)
~,
\label{eq:FX+}
\end{equation}
up to the factor coming from the integration over the zero-modes of 
$X^\pm$. 
It is convenient to define
\begin{equation}
\left\langle F\left[X^{+},X^{-}\right]\right\rangle_{\rho}
\equiv
\frac{
\left\langle 
F\left[X^{+},X^{-}\right]
\prod_{r=1}^{N}
        e^{-ip_{r}^{+}X^{-}}\left(Z_{r} ,\bar{Z}_{r}\right)
\right\rangle
}{
\left\langle 
\prod_{r=1}^{N}
        e^{-ip_{r}^{+}X^{-}}\left(Z_{r} ,\bar{Z}_{r}\right)
\right\rangle
}~,
\end{equation}
and we obtain
\begin{equation}
\left\langle F\left[X^{+}\right]\right\rangle _{\rho} 
 =  F\left[-\frac{i}{2}\left(\rho+\bar{\rho}\right)\right]\;.
\label{eq:expX+}
\end{equation}

\subsubsection*{
$F\left[-\frac{i}{2}\left(\rho+\bar{\rho}\right)\right]
\exp
\left(-\frac{d-26}{24}\Gamma 
  \left[\ln\left(\partial \rho \bar{\partial}\bar{\rho}\right)\right]
\right)$ as a generating functional}

The manipulation to obtain eq.(\ref{eq:FX+}) is rather formal because 
$\partial\rho (z)$ possesses zeros and poles and 
we need to specify the regularization procedure to define 
$\Gamma 
\left[\ln\left(\partial \rho \bar{\partial}\bar{\rho}\right)\right]$. 
As we mentioned in the last section, 
this was done in Ref.~\cite{Mandelstam:1985ww} and 
$\Gamma \left[\ln \left(\partial \rho \bar{\partial}\bar{\rho}
                  \right)\right]$ 
is given for arbitrary Mandelstam mapping $\rho$. 
%
Assuming that eq.(\ref{eq:FX+}) is true with $\Gamma$
derived in Ref.~\cite{Mandelstam:1985ww}
(and also in appendix \ref{sec:Gamma}), 
one can calculate correlation functions which involve $X^-$. 
Roughly speaking, 
differentiating the left hand side of eq.(\ref{eq:FX+}) 
with respect to $p_N^+$ and setting $p_N^+=0$, we obtain
\begin{equation}
\left\langle 
F\left[X^{+}\right]X^-(Z_N,\bar{Z}_N)
\prod_{r=1}^{N-1}
        e^{-ip_{r}^{+}X^{-}}\left(Z_{r} ,\bar{Z}_{r}\right)
\right\rangle
~,
\end{equation}
although the momentum conservation condition complicates 
the procedure a bit. 
In the same way, we can in principle obtain arbitrary 
correlation functions, 
treating the right hand side of eq.(\ref{eq:FX+}) as a 
kind of generating functional. 

Since all the correlation functions on the complex plane 
are given this way, 
we define the theory based on eq.(\ref{eq:FX+}), 
rather than starting from the action (\ref{eq:SXpm}). 
If one define the theory in this way, 
it is not a priori clear if $T_{X^\pm}(z)$  
in eq.(\ref{eq:emtensorXpm}) can be considered as 
the energy-momentum tensor. 
In the rest of this section, we would like to calculate the correlation 
functions of $T_{X^\pm}(z)$ and examine 
if it can be regarded as the energy-momentum tensor of the theory. 

One comment is in order. 
The operator $e^{-ip^+X^-}$ can be considered to create a hole 
of length $\alpha$ in the light-cone diagram. 
Therefore it is similar to the macroscopic observables 
in the old matrix models \cite{Banks:1989df}. 
These operators are nonlocal objects and give rise to singularities at 
$z_I$ where 
no operators are inserted. 
Taking $p^+\to 0$ limit to obtain local operators is exactly 
what was done in the old matrix models.  

\subsection{Correlation functions of $X^{-}$}
\label{sec:correlation}

In order to calculate the correlation functions of 
$T_{X^\pm}(z)$, we need correlation functions of $X^-$. 
Let us follow the above-mentioned procedure and calculate 
them. 

\subsubsection*{One point function}

First we consider the simplest example,
$\left\langle \partial X^{-}\left(z\right)\right\rangle_{\rho}$.
In order to calculate this quantity we start from
\begin{equation}
\left\langle 
\prod_{r=0}^{N+1}
        e^{-ip_{r}^{+}X^{-}}\left(Z_{r} ,\bar{Z}_{r}\right)
\right\rangle
\sim
\exp
\left(-\frac{d-26}{24}\Gamma 
\left[\ln
\left(\partial \rho^\prime \bar{\partial}\bar{\rho}^\prime\right)
\right]
\right)
~,
\end{equation}
where $p_{N+1}^+=-p_0^+$ and
\begin{equation}
\rho' (z) = \sum_{r=0}^{N+1} \alpha_{r} \ln (z-Z_{r})~,
 \qquad \alpha_{N+1}=-\alpha_{0}~.
\end{equation}
Then the one point function
$\left\langle \partial X^{-}\left(z\right)\right\rangle_{\rho}$
 can be given as
\begin{equation}
\left\langle \partial X^{-}\left(Z_{0}\right)\right\rangle_{\rho}
 =  
       2i \left.
        \partial_{Z_{0}}\partial_{\alpha_{0}}
        \left(- \frac{d-26}{24} \Gamma 
        \left[\ln
        \left(\partial \rho^\prime \bar{\partial}\bar{\rho}^\prime
        \right)\right]
                  \right)
              \right|_{\alpha_{0}=0}
       \;.
\label{eq:expdX-}
\end{equation}

It is straightforward to evaluate 
the right hand side of eq.(\ref{eq:expdX-})
by using the expression~(\ref{eq:Gamma-c}) for $\Gamma [\phi ]$. 
In the calculations, one should notice the following points. 
While there are $N$ interaction points $z'_{I'}$ for $\rho'$,
there are only $N-2$ interaction points $z_{I}$
for $\rho=\lim_{\alpha_{0} \rightarrow 0} \rho'$.
Since
\begin{equation}
\partial \rho' (z)
 = \partial\rho (z)  + \frac{\alpha_{0}}{z-Z_{0}}
   -\frac{\alpha_{0}}{z-Z_{N+1}}~,
\label{eq:drhohat}
\end{equation}
what happens
in the limit $\alpha_{0} \rightarrow 0$ is 
as follows:
One of $z'_{I'}$'s,
       which we denote by $z'_{I^{(0)}}$,
goes to $Z_{0}$;
One of $z'_{I'}$'s, 
       which we denote by $z'_{I^{(N+1)}}$,
goes to $Z_{N+1}$;
The other interaction points $z'_{I}$ 
     go to the interaction points $z_{I}$ for $\rho$,
which are denoted with the same subscripts. 
As we mentioned earlier, $e^{-ip^{+}_0X^{-}}(Z_0)$ with finite $p_0^{+}$
should be considered as a nonlocal operator,
which induces singularities at the interaction points where there are 
no operator insertions. 
$z'_{I^{(0)}}\to Z_{0}$ in the limit $p_0^+\to 0$ is consistent with 
the fact that $e^{-ip^{+}_0X^{-}}(Z_0)$ tends to a local operator. 
One subtle point to notice is that if $Z_0$ is close to 
one of $Z_r~(r=1,\cdots ,N)$ 
and the interaction point nearest to $Z_r$ coincides with $z'_{I^{(0)}}$, 
the Neumann coefficient $\bar{N}_{\ 00}^{\prime rr}$ does not go to 
$\bar{N}_{00}^{rr}$ in the limit $p_0^+\to 0$. 
Since $\Gamma \left[\ln
\left(\partial \rho^\prime \bar{\partial}\bar{\rho}^\prime\right)
\right]$ depends on $\bar{N}_{\ 00}^{\prime rr}$,
it implies that $e^{-ip^{+}_0X^{-}}(Z_0)$ has a nonlocal effect even 
if we take $p_0^+\to 0$ in such a configuration. 
Then we cannot expect to get correlation functions of local operators 
from $\Gamma \left[\ln
\left(\partial \rho^\prime \bar{\partial}\bar{\rho}^\prime\right)
\right]$. 
In the following, 
we will assume that $Z_0$ and $Z_{N+1}$ are not close to 
any of $Z_r~(r=1,\cdots ,N)$, 
to avoid such a situation. 
Namely we assume that $\rho (Z_0)$ and $\rho (Z_{N+1})$ are not 
in the regions of external propagators in the $\rho$ plane. 
We calculate 
$\left\langle \partial X^{-}\left(Z_0\right)\right\rangle _{\rho}$ 
for such $Z_0$, 
and analytically continue the result to the whole complex 
plane.

Using the identities presented in appendix \ref{sec:relations},
we obtain
\begin{eqnarray}
\left\langle \partial X^{-}\left(z\right)\right\rangle _{\rho}
 & = &  \frac{d-26}{24}
       \, 2i\left[
       -
        \sum_{r=1}^{N} \frac{1}{\alpha_{r}}
            \left( \frac{1}{z-z_{I}^{(r)}} - \frac{1}{z-Z_{r}}
            \right)
         \right.
        \nonumber\\
  && \hspace{4.5em}  
     {} - \sum_{I} \frac{1}{\partial^{2} \rho (z_{I})}
         \frac{1}{(z-z_{I})^{2}}
             \frac{\partial (-W)}{\partial z_{I}}
        \nonumber\\
  && \hspace{4.5em} \left.
     -3 \sum_{I}
     \left(
      \frac{1}{\partial^{2} \rho (z_{I})}
         \frac{1}{(z-z_{I})^{3}}
      -\frac{\partial^3\rho (z_I)}
        {2\left(\partial^2\rho (z_I)\right)^2}\frac{1}{(z - z_I)^2}
      \right)
     \right],~~~
\label{eq:dX-}
\end{eqnarray}
where $W$ is defined in eq.(\ref{eq:W}).

One can generalize eq.(\ref{eq:expdX-})
and calculate the correlation functions with insertions of $X^+$ as
\begin{eqnarray}
& &
\left\langle
   \partial X^{-}\left(Z_{0}\right)  F \left[X^{+}\right]
\right\rangle _{\rho}
\nonumber
\\
& &
\hspace{5mm}
= 
        \left. 
\frac{
2i
\partial_{Z_0}
\partial_{\alpha_0}
\left\langle 
F\left[X^{+}\right]
\prod_{r=0}^{N+1}
        e^{-ip_{r}^{+}X^{-}}\left(Z_{r} ,\bar{Z}_{r}\right)
\right\rangle
}
{
\left\langle 
\prod_{r=0}^{N+1}
        e^{-ip_{r}^{+}X^{-}}\left(Z_{r} ,\bar{Z}_{r}\right)
\right\rangle
}                
         \right|_{\alpha_{0}=0}   
\nonumber \\
 & &
 \hspace{5mm}
 =
    \left\langle 
        \partial X^{-}\left(Z_{0}\right)
    \right\rangle _{\rho}
    F \left[ -\frac{i}{2} \left( \rho+\bar{\rho} \right) \right]
  + \int d^{2}z \, \frac{1}{Z_{0}-z}
    \left. \frac{\delta F\left[X^{+}\right]}
                {\delta X^{+} (z,\bar{z})}
    \right|_{X^{+}=-\frac{i}{2}\left(\rho+\bar{\rho}\right)}\;.
\label{eq:expXpm}
\end{eqnarray}
Since we are dealing with the correlation functions with 
source terms for $X^-$, 
we can read off the operator relations from these correlation functions. 
{}From eqs.(\ref{eq:expX+}) and (\ref{eq:expXpm}), 
we obtain the OPE's
\begin{equation}
\partial X^{+} (z)
\partial X^{+} (z')
 \sim \mathrm{regular}~,
\qquad
\partial X^{-} (z)
\partial X^{+} (z')
 \sim \frac{1}{(z-z')^2}~,
\label{eq:OPE-Xpm}
\end{equation}
which are valid if $z$ and $z'$ are away from the singularities
$Z_r,z_I$. 
These are consistent with eq.(\ref{eq:OPE}).

\subsubsection*{Two point function}
The two point function for $\partial X^-$ can be calculated
by using eq.(\ref{eq:dX-}) as
\begin{equation}
\left\langle
   \partial X^{-} (z)\partial X^{-} (Z_0)
\right\rangle _{\rho} 
=
\left.
2i\partial_{Z_0}\partial_{\alpha_0}
\left\langle \partial X^{-}\left(z\right)\right\rangle _{\rho^\prime}
\right|_{\alpha_0=0}
+
\left\langle \partial X^{-}\left(z\right)\right\rangle _{\rho}
\left\langle \partial X^{-}\left(Z_0\right)\right\rangle _{\rho}
~.
\label{eq:twopoint}
\end{equation}
Here we are interested in the singularity at $z=Z_0$. 
It is straightforward to calculate the right hand side 
of eq.(\ref{eq:twopoint}) and obtain
\begin{equation}
\left\langle
   \partial X^{-} (z)\partial X^{-} (Z_0)
\right\rangle _{\rho} 
 =  \left(2i\right)^{2}
    \left( -\frac{d-26}{12} \right)
\partial_{z} \partial_{Z_0} \left[
  \frac{1}{(z-Z_0)^{2}}
  \frac{1}{\partial\rho (z)\partial\rho (Z_0)}
  \right] 
+\mbox{regular terms}\;.
\label{eq:dX-2}
\end{equation}
{}From this, we deduce the OPE
\begin{eqnarray}
\partial X^{-} (z) \partial X^{-} (z')
 &\sim& -\frac{d-26}{12}
        \partial_{z} \partial_{z'}
        \left[ \frac{1}{(z-z')^{2}}
               \frac{1}{\partial X^{+}(z) \partial X^{+}(z')}
        \right]
 \nonumber\\
 &\sim&  {} -\frac{d-26}{12}
   \left[- \frac{1}{(z-z')^{4}}
          \frac{6}{\left(\partial X^{+}(z')\right)^{2}}
        - \frac{1}{(z-z')^{3}}
          3 \partial \left( 
                \frac{1}{\left( \partial X^{+} (z') \right)^{2}}
               \right)
   \right. \nonumber\\
 && \hspace{5em} \left.
        {}-\frac{1}{(z-z')^{2}} \frac{1}{2}
          \partial^{2} \left(
                 \frac{1}{\left( \partial X^{+} (z') \right)^{2}}
                 \right)
   \right]~,
\label{eq:OPE3}
\end{eqnarray}
which is valid if $z$ and $z'$ are away from 
the singularities $Z_r,z_I$. 
This is consistent with eq.(\ref{eq:OPE}). 

We can also obtain an expression for
the correlation functions with $X^+$ insertions 
as we did in eq.(\ref{eq:expXpm}):
\begin{eqnarray}
\lefteqn{
  \left\langle \partial X^{-}\left(z\right)
               \partial X^{-}\left(w\right)
               F\left[X^{+}\right]
  \right\rangle _{\rho}
} \nonumber\\
 &=& \left\langle \partial X^{-}\left(z\right)
            \partial X^{-}\left(w\right)\right\rangle _{\rho}
      F\left[-\frac{i}{2}\left(\rho+\bar{\rho}\right)\right]
\nonumber \\
 &  & \quad {}+\left\langle \partial X^{-}\left(z\right)
               \right\rangle _{\rho}
       \int d^{2}w' \, \frac{1}{w-w'}
           \left.\frac{\delta F\left[X^{+}\right]}
                      {\delta X^{+} (w',\bar{w}^\prime )}
           \right|_{X^{+}=-\frac{i}{2}\left(\rho+\bar{\rho}\right)}
\nonumber \\
 &  & \quad {}
   +\left\langle \partial X^{-}\left(w\right)
               \right\rangle _{\rho}
       \int d^{2}z' \, \frac{1}{z-z'}
            \left.\frac{\delta F\left[X^{+}\right]}
                       {\delta X^{+} (z',\bar{z}^\prime )}
            \right|_{X^{+}=-\frac{i}{2}\left(\rho+\bar{\rho}\right)}
\nonumber \\
 &  & \quad {}+\int d^{2}z' \int d^{2}w' \,
     \frac{1}{z-z'}  \frac{1}{w-w'}
     \left.\frac{\delta^{2}F\left[X^{+}\right]}
                {\delta X^{+}(z',\bar{z}')
                   \delta X^{+} (w',\bar{w}')}
     \right|_{X^{+}=-\frac{i}{2}\left(\rho+\bar{\rho}\right)}\;.
\label{eq:expdX-dX-F}
\end{eqnarray}

\subsection{Energy-momentum tensor}
\label{sec:propTpm}

To be precise, the term
$\partial X^{+}\partial X^{-}(z)$
in  $T_{X^\pm}(z)$ given in eq.(\ref{eq:emtensorXpm})
is defined as
\begin{equation}
: \! \partial X^{+} \partial X^{-} (z) \! : \;
 = \lim_{z'\rightarrow z}
    \left(  \partial X^{+} (z') \partial X^{-} (z)
             - \frac{1}{(z'-z)^{2}}
    \right)~.
\label{eq:subtraction-Xpm}
\end{equation}
Then the correlation functions with one $T_{X^\pm}(z)$ insertion 
can be evaluated as
\begin{eqnarray}
\lefteqn{
\left\langle T_{X^\pm} (z) F\left[X^{+}\right]\right\rangle_{\rho}
}\nonumber\\
  && =  \left\langle T_{X\pm} (z) \right\rangle_{\rho}
      F \left[ -\frac{i}{2}\left(\rho+\bar{\rho}\right) \right]
-\frac{i}{2}
          \partial\rho\left(z\right)
       \int d^{2}z' \, \frac{1}{z-z'}
         \left.\frac{\delta F\left[X^{+}\right]}
                    {\delta X^{+} (z',\bar{z}')}
         \right|_{X^{+}=-\frac{i}{2}\left(\rho+\bar{\rho}\right)},~~~
\label{eq:expTF}
\end{eqnarray}
where
\begin{equation}
\left\langle T_{X^\pm} (z) \right\rangle_{\rho}
= {} -\frac{i}{2} \partial\rho\left(z\right)
          \left\langle \partial X^{-}\left(z\right)
          \right\rangle _{\rho}
         -\frac{d-26}{12}\left\{ \rho,z\right\}~.
\label{eq:vevTpm}
\end{equation}

One can evaluate the 
right hand side of eq.(\ref{eq:vevTpm}) and 
examine how it behaves around the possible singularities:
\begin{equation}
\left\langle T_{X^\pm} (z) \right\rangle_{\rho}
\sim \left\{
  \begin{array}{ll}
     \displaystyle 
       \frac{1}{z-Z_{r}}
       \frac{\partial \left( - \frac{d-26}{24}\Gamma [\phi ] \right)}
            {\partial Z_{r}}
     & z\sim Z_{r}
    \\[1ex]
     \displaystyle
       \mathrm{regular} & z\sim z_{I} 
     \\[0.5ex]
       \mathcal{O} \left(\frac{1}{z^{4}} \right)
       & z\sim \infty
  \end{array}
\right. .
\label{eq:vevTpm2}
\end{equation}
Eqs.(\ref{eq:vevTpm2}) and (\ref{eq:expTF}) imply 
that $T_{X^\pm} (z)$ is regular at $z=z_{I}$ and $\infty$,
if no operators are inserted there. 
Therefore, although $X^-(z)$ is singular at $z=z_I$ without any operator 
insertions, the energy momentum tensor is regular and conserved. 
This property is essential for constructing the BRST charge.  
From eq.(\ref{eq:vevTpm2}) for $z\sim Z_{r}$ 
we can read off the OPE
\begin{equation}
T_{X^\pm}(z)
   e^{-ip_{r}^{+}X^{-}} (Z_{r},\bar{Z}_{r} ) 
\sim  
  \frac{1}{z-Z_{r}}\partial
      e^{-ip_{r}^{+}X^{-}} (Z_{r},\bar{Z}_{r})\;.
\label{eq:scalar}
\end{equation}
Although 
$e^{-ip_{r}^{+}X^{-}}$ is a nonlocal operator, 
it behaves as a primary field of weight $0$. 

Using the OPE's (\ref{eq:OPE-Xpm})
and (\ref{eq:OPE3}),
we can show that $T_{X^\pm}(z)$ satisfies
\begin{equation}
T_{X^\pm}(z) T_{X^\pm}(z')
\sim \frac{\frac{1}{2} (28-d)}{(z-z')^{4}}
     + \frac{2}{(z-z')^{2}} T_{X^\pm}(z')
     + \frac{1}{z-z'} \partial T_{X^\pm} (z')~.
\end{equation}
Therefore
the central charge of the Virasoro algebra
in the $X^{\pm}$ CFT is $28-d$.
We can also find that
\begin{equation}
T_{X^\pm} (z) \partial X^{\pm} (z')
\sim \frac{1}{(z-z')^{2}} \partial X^{\pm} (z')
   + \frac{1}{z-z'} \partial^{2} X^{\pm} (z')~,
\end{equation}
and thus $\partial X^{\pm}$ are primary fields of weight $1$.

\section{BRST Invariant Formulation in Noncritical Dimensions}
\label{sec:BRST}
Since the worldsheet theory for $X^\pm$ is a CFT with 
Virasoro central charge $28-d$,
with the transverse coordinates $X^i$ added
the total central charge 
of the system for $X^\pm$ and $X^i$ is $26$. 
Therefore with ghosts $b$ and $c$, 
we can construct a nilpotent BRST charge $Q_\mathrm{B}$. 

As we have shown in section \ref{sec:LCcov}, 
the amplitude for the light-cone gauge string field theory 
can be rewritten by using these variables.
We start from the correlation function given in eq.(\ref{eq:noncriticalF}), 
where the vertex operator $V_r^\mathrm{LC}$ is of the form (\ref{eq:VrLC}) 
but with the on-shell condition
\begin{equation}
\frac{1}{2} \left( -2 p^+_r p^-_r + p_r^ip_r^i \right)
+\mathcal{N}_{r}
=
\frac{d-2}{24}~.
\label{eq:nonshell}
\end{equation}
We can proceed in the same way as in section \ref{sec:LCcov} and 
eventually obtain
\begin{eqnarray}
 F
 &\sim & \int\left[dX^{\pm}dX^{i}dbdcd\tilde{b}d\tilde{c}\right]
 e^{-S_{X^{i}}-S_{X^{\pm}}-S_{bc}}
 \nonumber\\
 &  & \hspace{1cm}
 \times
 \prod_{I}
 \left(
 \oint_{C_{I}}\frac{dz}{2\pi i}\frac{b}{\partial\rho}(z)
 \oint_{C_{I}}
 \frac{d\bar{z}}{2\pi i}\frac{\tilde{b}}{\bar{\partial}\bar{\rho}}
  (\bar{z})
 \right)
 \nonumber
 \\
 & &
 \hspace{1cm}
 \times
 \prod_{r=1}^{N} \left(c\tilde{c}V_{r}^{\mathrm{DDF}}
 \exp \left(-i\frac{d-26}{24}\frac{X^+}{p_r^+}\right)
 \right)
 \left(Z_{r},\bar{Z}_{r}\right)
 \nonumber
 \\
 & &
 \hspace{1cm}
 \times
 \prod_{r=1}^{N}
 \exp \left(i\frac{d-26}{24}\frac{X^+}{p_r^+}\right)
 \left( z_I^{(r)},\bar{z}_I^{(r)} \right)~,
\label{eq:BRSTinv}
\end{eqnarray}
where $V_r^\mathrm{DDF}$ is defined in the same way as in 
eq.(\ref{eq:VDDF}). 

It is easy to show that 
$V_{r}^{\mathrm{DDF}}
 \exp \left(-i\frac{d-26}{24}\frac{X^+}{p_r^+}\right)$ 
is a primary field of weight $(1,1)$ and
$T_{X^\pm}(z)$ is regular 
even with the insertion 
$\exp \left(i\frac{d-26}{24}\frac{X^+}{p_r^+}\right)
  (z_I^{(r)},\bar{z}_I^{(r)})$. 
Therefore eq.(\ref{eq:BRSTinv}) gives a BRST invariant expression 
for the amplitude. 
$V_{r}^{\mathrm{DDF}}
 \exp \left(-i\frac{d-26}{24}\frac{X^+}{p_r^+}\right)$ 
may look as a vertex operator with momentum 
\begin{equation}
p_r^{\prime -}
=
p_r^-+\frac{1}{p_r^+}\frac{d-26}{24}~,
\end{equation}
instead of $p_r^-$, 
but the momentum which is conserved is $p^-$. 
The conserved momentum can be identified with the operator
\begin{equation}
\oint\frac{dz}{2\pi i}i\partial X^-(z)~,
\end{equation}
on the worldsheet, which is conserved at the interaction points with 
insertions 
$\exp \left(i\frac{d-26}{24}\frac{X^+}{p_r^+}\right)$.

\section{Discussions}
\label{sec:discussions}

In this paper, we have constructed a BRST invariant worldsheet
theory which corresponds to the light-cone 
gauge string field theory in $d~(d\neq 26)$ space-time dimensions.
The worldsheet theory for the longitudinal coordinate variables $X^\pm$ 
is different from the usual free theory, 
but it is a CFT with $c=28-d$. 
Our results provide yet another way to construct string theories 
in noncritical dimensions. 
The BRST invariant formulation will be useful to study 
D-branes for such string theories. 

Now that the CFT is given, we can at least formally construct
the interaction vertices of the string field theory
based on this CFT through the prescription of
Ref.~\cite{LeClair:1988sp}.
Since we have constructed the CFT
on the worldsheet of the light-cone string diagram,
the gauge unfixed version of the string field theory
is supposed to possess the joining-splitting type of interactions.
Such a theory looks similar to 
the $\alpha=p^{+}$ HIKKO theory given in Ref.~\cite{Kugo:1992md}.

The results in this paper should be generalized to be used 
in regularizing string field theory. 
One should consider the $X^\pm$ on the Riemann surfaces with 
higher genera, in order to check if it works for 
regularizing the UV and IR divergences. 
One should also construct a supersymmetric version of the CFT. 
The light-cone gauge 
superstring field theory in noncritical dimensions
can be used to  dimensionally regularize the tree amplitudes of 
the critical theory, as was discussed \cite{Baba:2009kr}. 
It is possible to generalize the calculations performed in this paper 
into the superstring case, 
although they are much more complicated. 
We will present these results elsewhere.

\section*{Acknowledgements}

This work was supported in part by 
Grant-in-Aid for Scientific Research~(C) (20540247)
and
Grant-in-Aid for Young Scientists~(B) (19740164) from
the Ministry of Education, Culture, Sports, Science and
Technology (MEXT).

%
%
\appendix

\section{Action of Light-Cone Gauge String Field Theory}
\label{sec:action}

In order to fix the notation,
we present the action for the light-cone gauge bosonic
string field theory in  $d$ space-time
dimensions.
The worldsheet variables are $X^i~(i=1,\cdots ,d-2)$. 
The action of the string field theory takes the form
\begin{eqnarray}
&&
S = \int dt \left[ \frac{1}{2} \int d1d2 
     \left\langle R(1,2) \left| \Phi \right\rangle_{1} \right.
       \left( i\frac{\partial}{\partial t}
              - \frac{L_{0}^{\mathrm{LC} (2)} 
              + \tilde{L}_{0}^{\mathrm{LC} (2)}
                       -\frac{d-2}{12}}
                     {\alpha_{2}}
       \right) |\Phi \rangle_{2}
     \right.
 \nonumber\\
  && \hspace{5em} \left.
    {}+ \frac{2g}{3} \int d1d2d3
         \left\langle V_{3}(1,2,3) \right|
        \Phi\rangle_{1} | \Phi\rangle_{2}
        |\Phi\rangle_{3} \right]~.
\label{eq:SFTaction}
\end{eqnarray}
Here
$L^{\mathrm{LC} (r)}_{0}$ is the zero-mode
of the transverse Virasoro generators for the $r$-th string,
$g$ is the coupling constant,
and $dr$ denotes the integration measure for
the momentum zero-modes of the $r$-th string defined as
\begin{equation}
dr=\frac{\alpha_{r}d\alpha_{r}}{4\pi}
   \frac{d^{d-2}p_{r}}{(2\pi)^{d-2}}~,
\label{eq:measure-r}
\end{equation}
where $\alpha_{r}=2p_r^{+}$ is the string-length parameter
of the $r$-th string.
$\left\langle R(1,2) \right|$ is the reflector given by
\begin{eqnarray}
\left\langle R(1,2) \right| &=& 
  \delta(1,2)
  \frac{1}{\alpha_{1}}
 \, {}_{2}\langle 0|\,{}_{1}\langle 0|
 e^{-\sum_{n=1}^{\infty} \frac{1}{n} \left(
     \alpha^{i(1)}_{n} \alpha^{i(2)}_{n}
    +\tilde{\alpha}^{i(1)}_{n} \tilde{\alpha}^{i(2)}_{n}
     \right)}~,
\nonumber\\
\delta (1,2)
&=&4\pi \delta \left( \alpha_{1}+\alpha_{2} \right)
  (2\pi)^{d-2}
  \delta^{d-2} \left(p_{1}+p_{2} \right)~.
\end{eqnarray}
$\left\langle V_{3} (1,2,3) \right|$ denotes
the three-string interaction vertex defined as 
\begin{eqnarray}
\left\langle V_{3} (1,2,3) \right|
&=& 4\pi \delta \biggl( \sum_{r=1}^{3} \alpha_{r} \biggr)
   \mathop{\mathrm{sgn}}(\alpha_{1}\alpha_{2}\alpha_{3})
     \left| 
        \frac{e^{-2\hat{\tau}_{0}\sum_{r=1}^{3}\frac{1}
             {\alpha_{r}}}}{\alpha_{1}\alpha_{2}\alpha_{3}}
     \right|^{\frac{d-2}{24}}
   \left\langle V_{3}^{\mathrm{LPP}} (1,2,3) \right|
  ~,
\end{eqnarray}
Here $\left\langle V^{\mathrm{LPP}}_{3} (1,2,3)\right|$
is the three-string LPP vertex \cite{LeClair:1988sp} 
for the transverse coordinate 
variables $X^i$ and 
$\hat{\tau}_{0}
=
\sum_{r=1}^{3} \alpha_{r} \ln |\alpha_{r}|$. 
The string field $|\Phi\rangle$ is taken to obey
the reality and the level-matching conditions. 

\section{Mandelstam Mapping}
\label{sec:Mandelstam}

Let $\rho$ be the standard complex coordinate on the $N$-string
tree diagram with the joining-splitting type interaction.
The portion on the $\rho$-plane corresponding to the $r$-th
external string $(r=1,\ldots,N)$ is mapped to the unit disk,
$|w_{r}| \leq 1$, of the $r$-th string as
\begin{equation}
\rho = \alpha_{r} \ln w_{r} + \tau^{(r)}_{0}+i\beta_{r}~,
\label{eq:rho-wr}
\end{equation}
where 
$\tau^{(r)}_{0}+i\beta_{r}$ is the coordinate on
the $\rho$-plane at which the $r$-th string interacts.

The $N$-string tree diagram is mapped to the complex $z$-plane
with $N$ punctures by
the Mandelstam mapping \cite{Mandelstam:1973jk}
\begin{equation}
\rho (z)=\sum_{r=1}^{N} \alpha_{r} \ln (z-Z_{r})~,
\qquad \sum_{r=1}^{N} \alpha_{r}=0~,
\label{eq:Mandelstam}
\end{equation}
where the puncture $z=Z_{r}$ correspond to the origin
of the unit disk $w_{r}=0$.
The $N-2$ interaction points $z_{I}~(I=1,\cdots ,N-2)$ are determined by
$\partial \rho (z_{I})=0$.
These are related to the interaction points on
the $\rho$-plane by
\begin{equation}
\rho ( z_{I}^{(r)} )
 = \tau^{(r)}_{0} + i\beta_{r}~,
\label{eq:rho-zIr}
\end{equation}
where $z_{I}^{(r)}$ denotes one of $z_{I}$
at which the $r$-th external string interacts.
The fact that $z_{I}$'s are the zeros of $\partial \rho (z)$
yields
\begin{eqnarray}
\partial \rho (z)
 &=& \left( \sum_{s=1}^{N} \alpha_{s} Z_{s} \right)
    \frac{\prod_{I} (z-z_{I})}
         {\prod_{r=1}^{N} (z-Z_{r})}~,
\label{eq:delrho2}
\\
\partial^{2} \rho (z_{I})
 &=& \left( \sum_{s=1}^{N} \alpha_{s} Z_{s} \right)
   \frac{\prod_{J \neq I} (z_{I} - z_{J})}
        {\prod_{r=1}^{N} (z_{I} - Z_{r}) }~,
\label{eq:deldelrho}
\\
\frac{\sum_{r=1}^{N} \alpha_{r} Z_{r}^{2}}
      { \sum_{s=1}^{N} \alpha_{s} Z_{s}}
&=& 
-\left( \sum_{I} z_{I}  - \sum_{r=1}^{N} Z_{r} \right)~,
\label{eq:zI-Zr}
\\
\alpha_{r}
 &=& \left( \sum_{s=1}^{N} \alpha_{s} Z_{s} \right)
   \frac{\prod_{I} (Z_{r} - z_{I})}
        { \prod_{s \neq r} (Z_{r} - Z_{s})}~.
\label{eq:Zr-Zs}
\end{eqnarray}

\section{Computation of $\Gamma [\phi ]$}
\label{sec:Gamma}

$\Gamma [\phi ]$ can be obtained by evaluating the Liouville action for
the metric (\ref{eq:metric}) on the 
$z$-plane~\cite{Mandelstam:1985ww}\cite{Green:1987mn}.
Here we will present an alternative derivation.
$e^{-\Gamma [\phi ]}$ can be regarded as the partition function 
of the light-cone gauge string theory in $26$ space-time dimensions
{}from the view point of the worldsheet CFT
on the light-cone diagram. 
Therefore the variation $\delta \Gamma [\phi ]$ under
\begin{equation}
Z_{r} \rightarrow Z_{r} + \delta Z_{r}~,
\quad
\bar{Z}_{r} \rightarrow \bar{Z}_{r} + \delta \bar{Z}_{r}~,
\quad
 \alpha_{r} \rightarrow \alpha_{r} + \delta \alpha_{r}
\qquad
(r=1,\ldots, N)~,
\label{eq:deltaZr}
\end{equation}
can be given by using the expectation value 
of the transverse energy-momentum tensor $T_{X^{i}}$. 
We can obtain $\Gamma [\phi]$ by integrating $\delta \Gamma [\phi]$. 
The result should be up to a factor 
which does not change under the variation eq.(\ref{eq:deltaZr}). 
Such a factor can be fixed by imposing the factorization condition 
in each case \cite{Baba:2009kr}. 

Before we begin the calculation, two comments are in order. 
Because of the constraint $\sum_{r=1}^{N} \alpha_{r} =0$,
all the $\delta \alpha_{r}$'s cannot be treated as independent
variations. Here we think of $\delta \alpha_{r}$ with $r=1,\cdots,N-1$
as independent of each other and $\delta \alpha_{N}$
as being determined by the relation
\begin{equation}
\delta \alpha_{N}=-\sum_{r=1}^{N-1} \delta \alpha_{r}~.
\label{eq:delta-alphaN}
\end{equation}
We also note that
$\Gamma [\phi]$ involves divergence originating from the infinite 
length of the external lines. 
We regularize it by cutting off the $r$-th external line 
so that its length becomes $\alpha_r\ell_r$, 
where $\ell_r$ is a large positive constant. 
By doing so, the contribution from the $r$-th external 
line to the partition function becomes $e^{-2\ell_r}$, 
and the divergences of $\Gamma [\phi]$ become 
independent of the parameters $\alpha_r, Z_r$. 
On the $z$-plane, this corresponds to cutting a hole 
of radius $\epsilon_r$ around $Z_r$, 
where $\epsilon_r$ and $\ell_r$ are related as
\begin{equation}
\alpha_{r}\ell_{r}
=
\mathrm{Re}
\left( 
\rho (z^{(r)}_{I}) - \rho (Z_{r}+\epsilon_{r}) 
\right)~.
\end{equation}

\subsubsection*{Variation $\delta \left(-\Gamma [\phi ]\right)$}
\begin{figure}[htbp]
\begin{center}
	\includegraphics[width=26.5em]{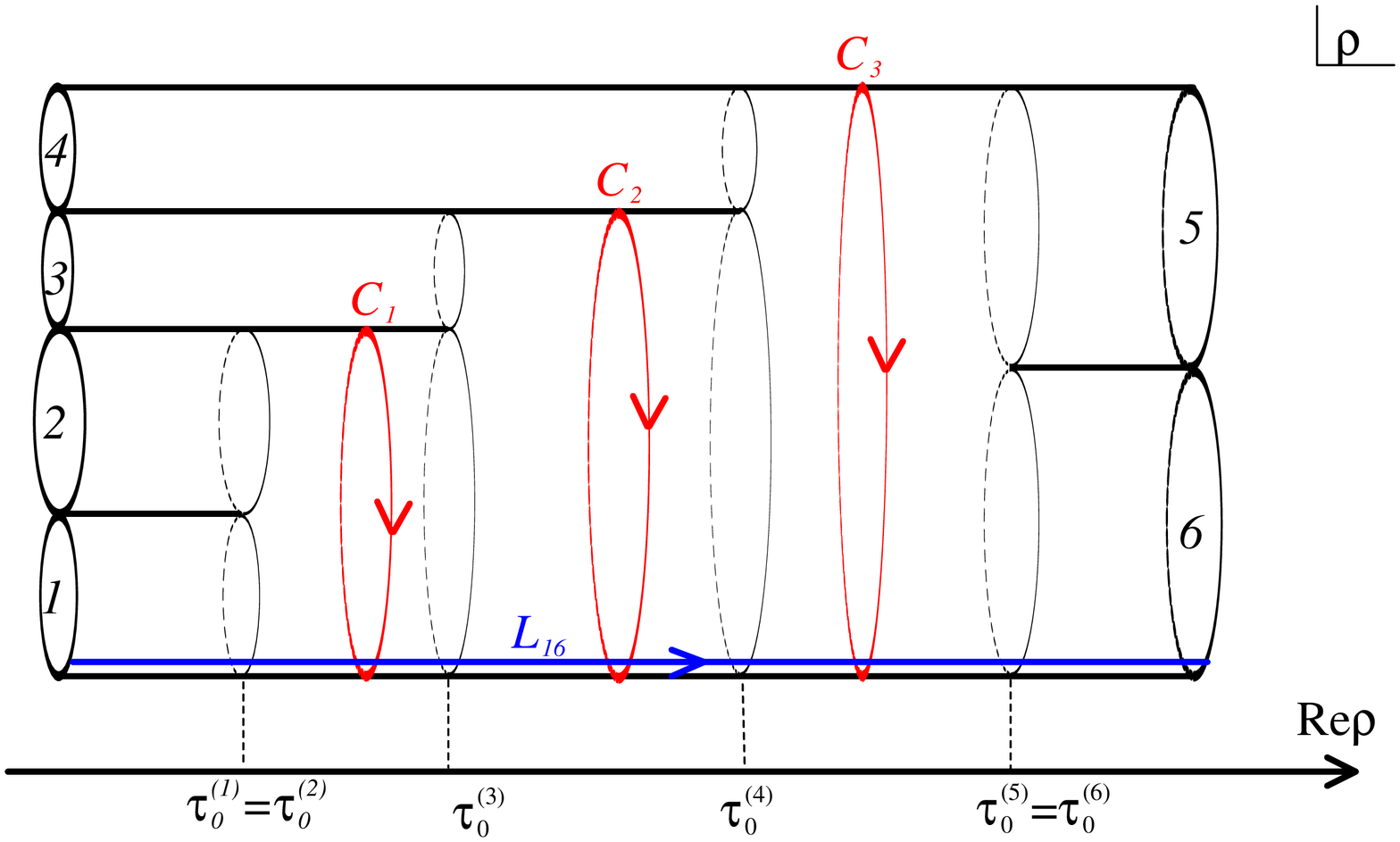}
	\caption{A typical $N$-string tree diagram with $N=6$.
          The contours $C_{I}$
          and the path $L_{1N}$ on the $\rho$-plane
          of the integrals in eq.(\ref{eq:variation1})
          are depicted for this case.}
	\label{fig:Npt}
\end{center}
\end{figure}

The variations (\ref{eq:deltaZr}) correspond to 
the variations of the
following parameters of the light-cone diagram:
$(\mathrm{I})$ the moduli parameters
\begin{equation}
\mathcal{T}_{I}
\equiv
\rho \left(z_{I+1}\right)
-
\rho \left(z_I\right)~;
\label{eq:mathcalT}
\end{equation}
$(\mathrm{II})$ the heights $\alpha_{r} \ell_{r}$
              of the external cylinders;
$(\mathrm{III})$ the circumferences $2\pi \alpha_{r}$
              of the external cylinders. 
Thus the variation $\delta (-\Gamma [\phi ])$ is expressed as
\begin{equation}
\delta \left( -\Gamma [\phi ] \right)
= (\mathrm{I}) + (\mathrm{II}) + (\mathrm{III})~,
\label{eq:variation0}
\end{equation}
where
\begin{eqnarray}
(\mathrm{I})
 &=&  \sum_{I}
   \delta \mathcal{T}_{I} \oint_{C_{I}} 
   \frac{d\rho}{2\pi i}
   \left\langle T_{X^{i}} (\rho) \right\rangle
 + \mathrm{c.c.}~,
\nonumber\\
(\mathrm{II})
 &=& \sum_{r=1}^{N-1}
   \left[
       -\ell_{r} \delta \alpha_{r} \oint_{\rho (Z_{r})}
    \frac{d\rho}{2\pi i} \left\langle T_{X^{i}} (\rho)
    \right\rangle
   + \ell_{N} \delta \alpha_{r} \oint_{\rho (Z_{N})}
    \frac{d\rho}{2\pi i}
    \left\langle T_{X^{i}}(\rho)
    \right\rangle
   \right]
   + \mathrm{c.c.}~,
  \nonumber\\
 (\mathrm{III})
  &=& \sum_{r=1}^{N-1}
     i 2\pi \delta \alpha_{r} \int_{L_{rN}} \frac{d\rho}{2\pi i}
    \left\langle T_{X^{i}} (\rho) \right\rangle
    + \mathrm{c.c.}~.
\label{eq:variation1}
\end{eqnarray}
Here $\mathrm{c.c.}$ stands for the complex conjugate.
The integration contour $C_{I}$ of the term $(\mathrm{I})$ 
lies between the consecutive interaction points $\rho (z_{I+1})$ 
and $\rho (z_{I})$ as depicted in Fig.~\ref{fig:Npt}.
The integration path $L_{rN}$ of term $(\mathrm{III})$
in eq.(\ref{eq:variation1}) is a line stretching from
the asymptotic region of the $r$-th external string
to that of the $N$-th string on the $\rho$-plane.
As an example, the path $L_{1N}$ is depicted in Fig.~\ref{fig:Npt}.
On the $z$-plane, the path $L_{rN}$ becomes a segment
connecting the two punctures $Z_{r}$ and $Z_{N}$
with the orientation from $Z_{r}$ to $Z_{N}$.

The expectation value of $T_{X^{i}} (\rho)$ can be 
evaluated by going to the $z$-plane. 
Since
\begin{equation}
T_{X^{i}} (\rho)
 = \frac{1}{( \partial \rho (z) )^{2}}
  \left( T_{X^{i}} (z)
        - 2 \{\rho,z\}
  \right)~,
\qquad
 \left\langle T_{X^{i}} (z) \right\rangle =0~,
\end{equation}
we obtain
\begin{equation}
\left\langle T_{X^{i}} (\rho) \right\rangle
 = \frac{1}{\left(\partial \rho (z) \right)^{2}}
   \left( -2 \{ \rho,z\} \right)~.
\label{eq:vevT}
\end{equation}
Using eq.(\ref{eq:vevT}),
one can rearrange the term $(\mathrm{I})$ in eq.(\ref{eq:variation1})
as
\begin{eqnarray}
&&
 \sum_{I}
  \delta \mathcal{T}_{I} \oint_{C_{I}}
  \frac{d\rho}{2\pi i}
   \left\langle T_{X^{i}} (\rho) \right\rangle
 \nonumber\\
 &&= \sum_{I}
   \delta \left( \rho (z_{I+1}) -  \rho (z_{I})  \right)
    \oint_{C_{I}} \frac{dz}{2\pi i}
    \frac{(-2 \{\rho,z\})  }{\partial \rho (z)}
 \nonumber\\
&&= {} \sum_{r=1}^{N} 
\delta \rho (z^{(r)}_{I})
\oint_{Z_{r}} \frac{dz}{2\pi i}
    \frac{(- 2 \{\rho,z\})}{\partial \rho (z)}
+\sum_{I} 
 \delta \rho (z_{I})
 \oint_{z_{I}} \frac{dz}{2\pi i}
    \frac{(- 2 \{\rho,z\})}{\partial \rho (z)}
    \nonumber\\
&&
\hspace{1cm}
+\delta \rho (z^{\infty}_I)
\oint_{\infty} \frac{dz}{2\pi i}
    \frac{(- 2 \{\rho,z\})}{\partial \rho (z)}~,
\label{eq:I-2}
\end{eqnarray}
where
$z^{\infty}_{I}$ denotes the interaction point closest to
$z=\infty$.

{}From eq.(\ref{eq:delta-alphaN}), one can easily find that
term $(\mathrm{II})$ becomes
\begin{equation}
(\mathrm{II}) = - \sum_{r=1}^{N} \ell_{r} \delta \alpha_{r}
   \oint_{Z_{r}} \frac{dz}{2\pi i}
   \frac{(- 2 \{\rho,z\})}{\partial \rho (z)}
+ \mathrm{c.c.}~.
\label{eq:II-2}
\end{equation}

\begin{figure}[htbp]
\begin{center}
	\includegraphics[width=25em]{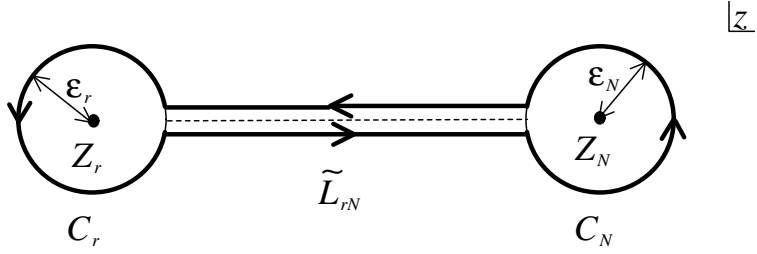}
	\caption{The integration path $\tilde{L}_{rN}$ consists
                 of the two oriented segments described by the
                 arrowed bold lines. The dashed line denotes
                 the cut of the integrand on the right hand side
                 of eq.(\ref{eq:log-int}).}
	\label{fig:cut}
\end{center}
\end{figure}

We can recast the $r$-th term of $(\mathrm{III})$
in eq.(\ref{eq:variation1}) into
\begin{eqnarray}
i2\pi \delta \alpha_{r} \int_{L_{rN}} \frac{d\rho}{2\pi i}
  \left\langle T_{X^{i}} (\rho) \right\rangle
&=&
i2\pi \delta \alpha_{r} \int_{L_{rN}} \frac{dz}{2\pi i}
 \frac{(-2 \{ \rho,z\})}{\partial \rho (z)}
\nonumber\\
&=& i2\pi \delta \alpha_{r} 
  \int_{\tilde{L}_{rN}} \frac{dz}{2\pi i}
   \frac{\ln (z-Z_{r}) - \ln (z-Z_{N})}{2\pi i}
   \frac{(-2 \{ \rho,z\})}{\partial \rho (z)}\;.~~~~~
\label{eq:log-int}
\end{eqnarray}
Here we have taken the cut of the function
$\ln (z-Z_{r}) - \ln (z-Z_{N})$ in the integrand on the right
hand side to be the segment $L_{rN}$ and the integration
path $\tilde{L}_{rN}$ to be the sum of the two oriented segments
connecting two punctures $Z_{r}$ and $Z_{N}$
as depicted in Fig.~\ref{fig:cut}.
Deforming the contours 
to rearrange eq.(\ref{eq:log-int}) further and
summing over $r$, we obtain
\begin{equation}
i2\pi \sum_{r=1}^{N}
  \delta \alpha_{r} \int_{L_{rN}} \frac{d\rho}{2\pi i}
  \left\langle T_{X^{i}} (\rho) \right\rangle
= - \left(
       \sum_{r=1}^{N} \int_{C_r} + \sum_{I} \oint_{z_{I}}
       + \oint_{\infty}
      \right)
     \frac{dz}{2\pi i} \delta \rho (z)
     \frac{(-2 \{ \rho,z\})}{\partial \rho (z)}~, 
\label{eq:III-2}
\end{equation}
where $C_r$ and $C_N$ are depicted in Fig. \ref{fig:cut}.

Eventually we obtain the expression
\begin{eqnarray}
\delta \left( -\Gamma [\phi ] \right)
&=& \left[ {}-\sum_{r=1}^{N} \int_{C_{r}} \frac{dz}{2\pi i}
     \frac{\delta \rho (z) - \delta \rho (z^{(r)}_{I})
            + \delta \alpha_{r} \ell_{r}}
          {\partial \rho (z)}
     \left(- 2  \{ \rho , z\} \right)
    \right.
\nonumber\\
&& \quad {}  - \sum_{I} \oint_{z_{I}} \frac{dz}{2\pi i}
     \frac{\delta \rho (z) - \delta \rho (z_{I})}
          {\partial \rho (z)}
     \left(- 2  \{ \rho , z\} \right)
\nonumber\\
&& \left.
   \quad {} - \oint_{\infty} \frac{dz}{2\pi i}
   \frac{\delta \rho (z) - \delta \rho (z^{\infty}_{I})}
        {\partial \rho (z)}
   \left( - 2 \{ \rho ,z\} \right)
   \right]
+\mathrm{c.c.}~.
\label{eq:variation2}
\end{eqnarray}

\subsubsection*{Evaluation of the contour integrals}
Let us evaluate the contour integrals on the right hand
side of eq.(\ref{eq:variation2}).
For $z\sim Z_r,z_I,\infty$, 
the Schwarzian derivative $\{ \rho , z\}$ behaves as
\begin{equation}
-2 \{\rho,z\}
 =
\left\{
 \begin{array}{ll}
 \displaystyle
     - \frac{1}{(z-Z_{r})^{2}}
     + \frac{1}{z-Z_{r}} \frac{\partial \left(-W\right)}{\partial Z_{r}}
     + \mathcal{O} \left( (z-Z_{r})^{0} \right)
& z\sim Z_{r}
\\[2ex]
\displaystyle
 \frac{3}{(z-z_{I})^{2}}
     + \frac{1}{z-z_{I}}
        \frac{\partial \left(-W\right)}{\partial z_{I}}
     + \mathcal{O} \left( (z-z_{I})^{0} \right)
& z \sim z_{I}
\\[2ex]
\mathcal{O} \left(\frac{1}{z^{4}} \right)
& z \sim \infty
\end{array}
\right.~,
\label{eq:schwarzian-around}
\end{equation}
where $W$ is a function of $Z_r,\bar{Z}_r,z_I,\bar{z}_I$ defined as
\begin{equation}
W(Z_r,\bar{Z}_r,z_I,\bar{z}_I) \equiv -2 \left(
    \sum_{I > J} \ln |z_{I} -z_{J}|^{2}
    + \sum_{r>s} \ln |Z_{r} - Z_{s}|^{2}
    - \sum_{I,r} \ln |Z_{r}-z_{I}|^{2}    \right)~.
\label{eq:W}
\end{equation}

For the contour integral along $C_r$, one should notice that 
$\delta \rho (z)$ involves a term $\delta \alpha_r\ln (z-Z_r)$. 
This part can be integrated as
\begin{eqnarray}
\int_{C_r}\frac{dz}{2\pi i}
\frac{\delta \alpha_r\ln (z-Z_r)}{\partial \rho (z)}
\left(- 2  \{ \rho , z\} \right)
&\sim&
-\frac{1}{\alpha_r}\int_{C_r}\frac{dz}{2\pi i}
\frac{\delta \alpha_r\ln (z-Z_r)}{z-Z_r}
\nonumber
\\
&=&
-\frac{\delta\alpha_r}{\alpha_r}
\int \frac{d\theta}{2\pi}
\left(\ln\epsilon_r +i\theta\right)
\nonumber
\\
&=&
-\frac{\delta\alpha_r}{\alpha_r}
\left(\ln\epsilon_r+\mathrm{imaginary~part}\right)~,
\end{eqnarray}
for $\epsilon_r \sim 0$. 
The other terms can be evaluated by taking the residues 
of the simple poles and one can show 
\begin{eqnarray}
& &
-\sum_{r=1}^{N} \int_{C_{r}} \frac{dz}{2\pi i}
     \frac{\delta \rho (z) - \delta \rho (z^{(r)}_{I})
            + \delta \alpha_{r} \ell_{r}}
          {\partial \rho (z)}
     \left(- 2  \{ \rho , z\} \right)+\mathrm{c.c}
\nonumber
\\
& &
\hspace{1cm}
=
\delta 
\left(
-2\sum_{r=1}^N
\mathrm{Re}\bar{N}^{rr}_{00}
\right)
+\sum_{r=1}^N 
\left(
\delta Z_r\frac{\partial \left(-W\right)}{\partial Z_r}
+
\mathrm{c.c.}
\right)~,
\end{eqnarray}
where $\bar{N}^{rr}_{00}$ is a Neumann coefficient
given by
\begin{equation}
\bar{N}^{rr}_{00}
 = - \sum_{s\neq r} \frac{\alpha_{s}}{\alpha_{r}}
   \ln (Z_{r} - Z_{s}) + \frac{\tau^{(r)}_{0}+i\beta_{r}}
                              {\alpha_{r}}~,
\label{eq:Nbarrr00}
\end{equation}
and $\tau^{(r)}_{0}+i \beta_{r}$ is defined
in eq.(\ref{eq:rho-zIr}).

For $z\sim z_I$, one can show
\begin{equation}
\frac{\delta \rho (z) - \delta \rho (z_{I})}{\partial \rho (z)}
 =  - \delta z_{I}
     + (z-z_{I}) \delta
       \left( \frac{1}{2} \ln \partial^{2} \rho (z_{I}) \right)
     + \mathcal{O} \left( (z-z_{I})^{2} \right)~,
\end{equation}
using
\begin{equation}
\delta z_{I}
 = - \frac{\partial \delta \rho (z_{I})}
          {\partial^{2} \rho (z_{I})}~,
\label{eq:delta-zI}
\end{equation}
which is derived by varying
the equation $\partial \rho (z_{I}) =0$
under eq.(\ref{eq:deltaZr}).
The contour integral around $z_I$ can be obtained as
\begin{eqnarray}
& &
- \sum_{I} \oint_{z_{I}} \frac{dz}{2\pi i}
     \frac{\delta \rho (z) - \delta \rho (z_{I})}
          {\partial \rho (z)}
     \left(- 2  \{ \rho , z\} \right)+\mathrm{c.c.}
\nonumber
\\
& &
\hspace{1cm}
=
\delta 
\left(
-3
\sum_I
\ln \left|\partial^2\rho (z_I)\right|
\right)
+\sum_I
\left(
\delta z_I\frac{\partial \left(-W\right)}{\partial z_I}
+
\mathrm{c.c.}
\right)~.
\end{eqnarray}

It is easy to see that the integral around $\infty$ vanishes.

Putting all the pieces together, 
we obtain
\begin{equation}
- \Gamma [\phi ]
= - W
    -2 \sum_{r=1}^{N}
       \mathop{\mathrm{Re}} \bar{N}^{rr}_{00} 
     -3\sum_{I}\ln 
      \left| \partial^{2} \rho (z_{I}) \right|
~.
\label{eq:Gamma-c}
\end{equation}
This form of $\Gamma [\phi ]$ is useful in the calculations in 
section \ref{sec:XpmCFT}. 

\subsubsection*{Other expressions}
We note that $W $ can be described as
\begin{equation}
W = 2\sum_{r=1}^{N} \ln \left|\alpha_{r} \right| 
         -4 \ln
         \left| \sum_{r=1}^{N} \alpha_{r} Z_{r}
         \right|
         -2 \sum_{I}
          \ln \left| \partial^{2} \rho (z_{I}) \right|~,
\label{eq:W2}
\end{equation}
which follows from eqs.(\ref{eq:deldelrho}) and (\ref{eq:Zr-Zs}).
This yields
\begin{equation}
e^{-\Gamma [\phi ]}
 =
  \prod_{r=1}^{N} \left| \alpha_{r}  \right|^{-2}
     \left| \sum_{s=1}^{N} \alpha_{s} Z_{s}
     \right|^{4}
     e^{-2 \sum_{r=1}^{N}
          \mathop{\mathrm{Re}} \bar{N}^{rr}_{00} }
    \prod_{I} \left| \partial^{2} \rho (z_{I})
              \right|^{-1}~.
\label{eq:exp-Gamma}
\end{equation}
This expression  is the one obtained by 
the method in Ref.~\cite{Mandelstam:1985ww}. 
We also note that
\begin{equation}
e^{-\Gamma [\phi ]}
 =
  \prod_{r=1}^{N} \left| \alpha_{r}  \right|^{-1}
  e^{-\frac{1}{2}W}
     \left| \sum_{s=1}^{N} \alpha_{s} Z_{s}
     \right|^{2}
     e^{-2 \sum_{r=1}^{N}
          \mathop{\mathrm{Re}} \bar{N}^{rr}_{00} }
    \prod_{I} \left| \partial^{2} \rho (z_{I})
              \right|^{-2}~,
\label{eq:exp-Gamma2}
\end{equation}
which is used 
in sections \ref{sec:LCcov} and \ref{sec:BRST}.



\section{Correlation
         Functions of $\partial X^{-}(z)$'s}
\label{sec:relations}

In this appendix, we present the details of the 
calculations to obtain
eqs.(\ref{eq:dX-}) and (\ref{eq:dX-2}).

Let us consider the interaction points
$z'_I, z'_{I^{(0)}}$ and $z'_{I^{(N+1)}}$ for
$\rho' (z)$ defined in eq.(\ref{eq:drhohat}),
which tend to $z_I,Z_{0}$ and $Z_{N+1}$
as $\alpha_{0} \rightarrow 0$ respectively.
{}From eq.(\ref{eq:drhohat}),
we can obtain the expansion of
$z'_I,z'_{I^{(0)}}$ and $z'_{I^{(N+1)}}$ 
in terms of $\alpha_{0}$, 
\begin{eqnarray}
&& z'_{I}-z_{I}
 =
 {}-\frac{\alpha_0}{\partial^2\rho\left(z_{I}\right)}
 \left(
 \frac{1}{z_I-Z_0}-\frac{1}{z_I-Z_{N+1}}
 \right)
 +  \mathcal{O} \left( \alpha_{0}^{2} \right)~,
\nonumber\\
&& z'_{I^{(0)}}-Z_{0}
 =
 {}-\frac{\alpha_0}{\partial\rho\left(Z_{0}\right)}
 \left[ 1 + 
    \left(
      \frac{\partial^{2}\rho\left(Z_{0}\right)}
           {\left(\partial\rho\left(Z_{0}\right)\right)^{2}}
      +\frac{1}{\partial\rho\left(Z_{0}\right)}
       \frac{1}{Z_{0}-Z_{N+1}}
    \right)
    \alpha_{0} 
 +  \mathcal{O} \left( \alpha_{0}^{2} \right) \right]~,
 \nonumber \\
&& z'_{I^{(N+1)}}-Z_{N+1}
\nonumber\\
&& \qquad \quad
 =
 \frac{\alpha_{0}}{\partial\rho\left(Z_{N+1}\right)}
 \left[1 -\left( \frac{\partial^{2}\rho\left(Z_{N+1}\right)}
              {\left(\partial\rho\left(Z_{N+1}\right)\right)^{2}}
        + \frac{1}{\partial\rho\left(Z_{N+1}\right)}
          \frac{1}{Z_{N+1}-Z_{0}}
  \right)
  \alpha_{0}
  +\mathcal{O} \left( \alpha_{0}^{2} \right)\right].~~~~~~~
\label{eq:ZI0-Z0}
\end{eqnarray}
The Neumann coefficients $\bar{N}^{\prime rr}_{\ 00} $
 for the Mandelstam mapping $\rho'$ behave as
\begin{eqnarray}
 \mathop{\mathrm{Re} } \bar{N}_{\ 00}^{\prime rr}
 & = &
  \mathop{\mathrm{Re} } \bar{N}_{00}^{rr}
  +  \frac{\alpha_0}{\alpha_{r}}
       \ln \left| \frac{\left(z_{I}^{(r)}-Z_{0}\right)
                          \left(Z_{r}-Z_{N+1}\right)}
                       {\left(Z_{r}-Z_{0}\right)
                           \left(z_{I}^{(r)}-Z_{N+1}\right)}
           \right|+  \mathcal{O} \left( \alpha_{0}^2\right)
  \qquad (r \neq 0,N+1)~,
\nonumber\\
     \mathop{\mathrm{Re}}  \bar{N}_{\ 00}^{\prime 00}
 & = & \ln \left|\frac{\alpha_0}{\partial\rho(Z_0)}\right|-1 
\nonumber \\
&&      + \mathop{\mathrm{Re}} \left(
            \frac{1}{2}
            \frac{\partial^{2}\rho\left(Z_{0}\right)}
               {\left(\partial\rho\left(Z_{0}\right)\right)^{2}}
      +
        \frac{1}{\partial\rho\left(Z_{0}\right)
                   \left(Z_{0}-Z_{N+1}\right)}
        \right)\alpha_0
       +\mathcal{O} \left( \alpha_{0}^2 \right) \;,
\nonumber\\
 \mathop{\mathrm{Re}} \bar{N}_{\ \;0  \ 0}^{\prime N+1N+1}
 & = & \ln \left|\frac{\alpha_0}{\partial\rho(Z_{N+1})}\right|-1 
\nonumber\\
&&      - \mathop{\mathrm{Re}} \left(
            \frac{1}{2}
            \frac{\partial^{2}\rho\left(Z_{N+1}\right)}
               {\left(\partial\rho\left(Z_{N+1}\right)\right)^{2}}
      +
        \frac{1}{\partial\rho\left(Z_{N+1}\right)
                   \left(Z_{N+1}-Z_{0}\right)}
        \right)\alpha_0
    +\mathcal{O} \left( \alpha_{0} ^2\right).~~~
\end{eqnarray}
Using eqs.(\ref{eq:delrho2}) and (\ref{eq:ZI0-Z0}),
we obtain
\begin{eqnarray}
 W^{\prime }
 &=& 8 \ln |\alpha_0| + W 
     -4\ln \left|\partial \rho(Z_0) \partial \rho(Z_{N+1})\right|
 \nonumber\\
&&{}   +2 \mathop{\mathrm{Re}} \left[
     -\sum_{I} \frac{1 }{\partial^{2} \rho (z_{I})}
               \left( \frac{1}{z_{I} -Z_{0}}
                      - \frac{1}{z_{I} - Z_{N+1}}
               \right)
               \frac{\partial W }{\partial z_{I}}
  \right.
  \nonumber\\
  && \qquad \qquad 
    {} +  4  
          \frac{\partial^{2} \rho (Z_{0})}
              {\left( \partial \rho (Z_{0}) \right)^{2}}
            - 4 \frac{\partial^{2} \rho (Z_{N+1})}
              {\left( \partial \rho (Z_{N+1}) \right)^{2}}
\nonumber\\
&& \qquad \qquad
\left.
{}     + 2 \frac{1}{\partial \rho (Z_{0})}
        \frac{1}{Z_{0}-Z_{N+1}} 
        - 2 \frac{1}{\partial \rho (Z_{N+1})}
        \frac{1}{Z_{N+1}-Z_{0}}
  \right]\alpha_0 + \mathcal{O}\left( \alpha_{0}^2 \right),
\end{eqnarray}
and 
\begin{eqnarray}
\partial^{2}
  \rho' \left( z'_{I}\right)
 & = & \partial^{2}\rho\left(z_{I}\right)
       +  \left(
           -\frac{1}{\left(z_{I}-Z_{0}\right)^{2}}
           +\frac{1}{\left(z_{I}-Z_{N+1}\right)^{2}}
       \right)\alpha_0
  \nonumber \\
 &  & {} - 
           \frac{\partial^3 \rho\left(z_I\right)}
               {\partial^{2}\rho\left(z_{I}\right)}
          \left(
             \frac{1}{z_{I}-Z_{0}}-\frac{1}{z_{I}-Z_{N+1}}
          \right)\alpha_0
      + \mathcal{O} \left(\alpha_{0} ^2\right) \;,
\nonumber\\
   \partial^{2} \rho'
           \left(z'_{I^{\left(0\right)}}\right)
 &=& -\frac{\left(\partial\rho\left(Z_{0}\right)\right)^{2}}
    {\alpha_{0}}
\nonumber\\
&& \ \times \left[ 1 
   -\left( \frac{3\partial^{2}\rho\left(Z_{0}\right)}
         {\left(\partial\rho\left(Z_{0}\right)\right)^{2}}
   + \frac{2}{\partial\rho\left(Z_{0}\right)\left(Z_{0}-Z_{N+1}\right)}
\right)  \alpha_0
   +\mathcal{O} \left(\alpha_{0}^2 \right)\right] \;,
\nonumber\\
   \partial^{2} \rho'
           \left(z'_{I^{\left(N+1\right)}}\right)
 &=& \frac{\left(\partial\rho\left(Z_{N+1}\right)\right)^{2}}
    {\alpha_{0}}
\nonumber\\
&& \ \times \left[ 1 
   +\left( \frac{3\partial^{2}\rho\left(Z_{N+1}\right)}
         {\left(\partial\rho\left(Z_{N+1}\right)\right)^{2}}
   + \frac{2}{\partial\rho\left(Z_{N+1}\right)\left(Z_{N+1}-Z_{0}\right)}
\right)  \alpha_0
   +\mathcal{O} \left(\alpha_{0}^2 \right)\right]~.~~~~~
\end{eqnarray}
Gathering all the relations obtained above,
we have 
\begin{eqnarray}
\lefteqn{
\Gamma [\ln\left(\partial\rho'\bar{\partial}\bar{\rho}'\right)] 
= 6 \ln |\alpha_0 | - 4 
+ \Gamma [\ln\left(\partial\rho\bar{\partial}\bar{\rho}\right)]
}\nonumber\\
&& \hspace{4em}{}+ 2 \mathop{\mathrm{Re}}
  \left[ \sum_{r=1}^N \frac{1}{\alpha_r} 
   \ln \left|\frac{\left(z_I^{(r) }- Z_0\right)
                     \left(Z_r- Z_{N+1}\right)}
               {\left(Z_r - Z_0\right)
                      \left(z_I^{(r) }- Z_{N+1}\right)} \right|\right.
\nonumber\\
&&\hspace{8em}{}  
     -\sum_{I} \frac{1 }{\partial^{2} \rho (z_{I})}
               \left( \frac{1}{z_{I} -Z_{0}}
                      - \frac{1}{z_{I} - Z_{N+1}}
               \right)
               \frac{\partial W }{\partial z_{I}}
\nonumber\\
&&\hspace{8em} {}  -\frac{3}{2}\sum_{I}
           \frac{1}{\partial^2 \rho(z_I)}  \left(
           \frac{1}{\left(z_{I}-Z_{0}\right)^{2}}
           -\frac{1}{\left(z_{I}-Z_{N+1}\right)^{2}}
       \right)
\nonumber\\
&& \hspace{8em}
 \left. {}-\frac{3}{2}\sum_{I}
           \frac{\partial^3 \rho\left(z_I\right)}
               {\left( \partial^{2}\rho\left(z_{I}\right) \right)^2}
          \left(
             \frac{1}{z_{I}-Z_{0}}-\frac{1}{z_{I}-Z_{N+1}}
          \right)
   \right]\alpha_0 + \mathcal{O}\left( \alpha_{0}^2 \right).~~~~
\end{eqnarray}
We can see that $\lim_{\alpha_0\to 0}\Gamma [\ln\left(\partial\rho'\bar{\partial}\bar{\rho}'\right)]$ 
is divergent and does not coincide with 
$\Gamma [\ln\left(\partial\rho\bar{\partial}\bar{\rho}\right)]$. 
This singularity can be avoided by modifying
$\Gamma \to \Gamma -\sum_r\left(3\ln |\alpha_r|-2\right)$, 
which corresponds to a renormalization of the operator 
$e^{-ip_r^+X^-}$. 
Such a renormalization is irrelevant to the calculation of 
the right hand side of eq.(\ref{eq:expdX-}) and 
we obtain
\begin{eqnarray}
\lefteqn{
\partial_{Z_0}\partial_{\alpha_0}
 \left. 
   \Gamma [\ln\left(\partial\rho'\bar{\partial}\bar{\rho}'\right)] 
 \right|_{\alpha_0=0}
 =\partial_{Z_0} \left[ \sum_{r=1}^N \frac{1}{\alpha_r} 
\ln \left(\frac{z_I^{(r) }- Z_0}    
               {Z_r - Z_0} \right)\right.
     -\sum_{I} \frac{1 }{\partial^{2} \rho (z_{I})}
                \frac{1}{z_{I} -Z_{0}}
               \frac{\partial W }{\partial z_{I}}
}\nonumber\\
&&{} \hspace{12em}  \left.-\frac{3}{2}\sum_{I}
           \frac{1}{\partial^2 \rho(z_I)} 
           \left( 
           \frac{1}{\left(z_{I}-Z_{0}\right)^{2}}
           + \frac{\partial^3 \rho\left(z_I\right)}
               { \partial^{2}\rho\left(z_{I}\right) }
             \frac{1}{z_{I}-Z_{0}}
           \right)
   \right].
\label{eq:dGamma}
\end{eqnarray}
From this equation, we can compute the right hand side of 
eq.(\ref{eq:expdX-})
and obtain eq.(\ref{eq:dX-}).


Let us evaluate the right hand side of eq.(\ref{eq:twopoint}).
This can be evaluated by using eq.(\ref{eq:dGamma})
with $Z_{0}$, $\rho(z_{I})$ and $I$
replaced by $z$, $\rho'(z'_{I'})$ and $I'$ respectively
and with the range of the index $r$ taken to be from
$0$ to $N+1$.
The terms in which we are interested are
the $r=0$ contribution of the first term
and the $I'=I^{(0)}$ case of the second and the third terms
in the square brackets on the right hand side of eq.(\ref{eq:dGamma})
with the replacements mentioned above.
In order to evaluate the first term, we use
\begin{eqnarray}
\lefteqn{
  \frac{1}{\alpha_{0}}
  \ln \frac{z-z'_{I^{\left(0\right)}}}{z-Z_{0}}
= \frac{1}{\partial\rho (Z_{0}) (z-Z_{0})}
}
\nonumber \\
 & & \qquad
   {} +
      \left[\frac{-\frac{1}{2}}
                 {\left(\partial\rho (Z_{0})\right)^{2}}
            \frac{1}{(z-Z_{0})^{2}}
       +\left(
           \frac{\partial^{2} \rho (Z_{0})}
                {\left(\partial\rho (Z_{0})\right)^{3}}
        +\frac{1}{\left(\partial\rho (Z_{0})\right)^{2}}
         \frac{1}{Z_{0}-Z_{N+1}}\right)\frac{1}{z-Z_{0}}
     \right]\alpha_{0}
\nonumber \\
 &  & \qquad
    {}+ \mathcal{O} \left( \alpha_{0}^{2} \right)\;,
\label{eq:for-rprime0}
\end{eqnarray}
which follows from eq.(\ref{eq:ZI0-Z0}).
For the computation of the second and the third terms,
we use
\begin{eqnarray}
\frac{1}{\partial^{2} \rho' \left( z'_{I^{(0)}} \right)}
&=& 
 {}-\frac{\alpha_{0}}
         {\left(\partial\rho (Z_{0})\right)^{2}}
  -  \left[ \frac{3 \partial^{3} \rho (Z_{0})}
                 {\left( \partial \rho (Z_{0}) \right)^{4}}
            + \frac{2}{\left( \partial \rho (Z_{0}) \right)^{3}}
              \frac{1}{Z_{0}-Z_{N+1}} \right]
     \alpha_{0}^{2}
  + \mathcal{O} \left( \alpha_{0}^{3} \right)\;,
\nonumber \\
\frac{1}{z-z'_{I^{(0)}}} 
& =&
   \frac{1}{z-Z_{0}}
   - \frac{1}{\partial\rho (Z_{0})}
    \frac{1}{ (z-Z_{0})^{2}} \alpha_{0}
   + \mathcal{O} \left( \alpha_{0}^{2} \right)\;,
\nonumber \\
\frac{\partial W^{\prime }}
     {\partial z'_{I^{(0)}}}
 &=&
 - \frac{\partial^3\rho'(z'_{I^{(0)}})}{\partial^2\rho'(z'_{I^{(0)}})}
 =- 2
   \partial \rho (Z_{0})
   \frac{1}{\alpha_{0}} 
  + \frac{2}{Z_{0}-Z_{N+1}}
  +\mathcal{O} (\alpha_{0})~.
\label{eq:for-I0}
\end{eqnarray}
Combining these relations, we obtain
eq.(\ref{eq:dX-2}).



\end{document}